\documentclass[aps,prl,tightenlines,twocolumn,superscriptaddress]{revtex4-2}
\usepackage{graphicx,color}
\usepackage{epstopdf}
\usepackage[utf8]{inputenc}
\usepackage{amsmath}
\usepackage{amsfonts}
\usepackage{amssymb}
\usepackage{physics}
\usepackage{subfig}
\usepackage[colorlinks=true,allcolors=blue]{hyperref}%
\begin{document}
\title{Weyl Semimetal Path to Valley Filtering in Graphene}
\author{Ahmed M. Khalifa}
\affiliation{Department of Physics and Astronomy, University of Kentucky, Lexington, KY 40506-0055}
\author{Ribhu K. Kaul}
\affiliation{Department of Physics and Astronomy, University of Kentucky, Lexington, KY 40506-0055}
\author{Efrat Shimshoni}
 \affiliation{Department of Physics, Bar-Ilan University,
  Ramat-Gan 52900, Israel}
\author{H.A. Fertig}
  \affiliation{Department
  of Physics, Indiana University, Bloomington, IN 47405}
\author{Ganpathy Murthy}
\affiliation{Department of Physics and Astronomy, University of Kentucky, Lexington, KY 40506-0055}

  \begin{abstract}
  We propose a device in which a sheet of graphene is coupled to a Weyl semimetal, allowing for the physical access to the study of tunneling from two-dimensional to three dimensional massless Dirac fermions.  Due to the reconstructed band structure, we find that this device acts  as a robust valley filter for electrons in the graphene sheet. We show that, by appropriate alignment, the Weyl semimetal draws away current in one of the two graphene valleys while allowing current in the other to pass unimpeded. 
  In contrast to other proposed valley filters, the mechanism of our proposed device occurs in the bulk of the graphene sheet, obviating the need for carefully shaped edges or dimensions.
  \end{abstract}
  
\maketitle

Weyl semimetals (WSMs)
\cite{Murakami2007,Wan2011,Yang2011,Burkov2011,Xu2011} are
three-dimensional materials with an even number of isolated band
touching points in the Brillouin zone (BZ) called Weyl nodes. The band
dispersion near each Weyl node is that of a massless Weyl fermion,
which is chiral, the chirality being encoded in the Berry flux pierced
by a surface in momentum space enclosing the Weyl node. Either
inversion \cite{Murakami2007} or time-reveral symmetry
\cite{Wan2011,Yang2011,Burkov2011,Xu2011} or both must be broken in
WSMs. Many examples of WSM materials are now known
\cite{Lv2015,Xu2015a,Xu2015b,Inoue2016,SilvaNeto2019,Belopolski2019}. In
particular, one recent material, Co$_3$Sn$_2$S$_2$,
\cite{Liu2019,Morali2019} breaks both time-reversal and inversion, and
possesses cleaved surfaces with three-fold symmetry.
        
In monolayer graphene (MLG) \cite{GrapheneReview} electrons near
charge neutrality belong to one of the two Dirac points (${\bf K}$ and
${\bf K}'$, related to each other by inversion and time-reversal)
which constitute valleys. Due to the large difference in lattice
momentum, the valley degree of freedom is highly conserved in
transport. This has made it a promising material for use in
valleytronics, which seeks to use the valley degree of freedom to
encode and manipulate information \cite{ValleytronicsReview}. Either
electrons or excitons can be used to encode information; in the
following we will focus on electrons. A necessary first step in this
valleytronics program in MLG is to be able to produce valley-polarized
current, usually done by valley-filtering an incident
valley-unpolarized current. There are many theoretical proposals for
doing so.  Methods preserving the time-reversal of MLG while breaking
inversion \cite{Xiao2007} include a constriction with tailored edges
\cite{Rycerz2007}, using the ``high-energy'' dispersion of electrons
away from the Dirac points \cite{trigonal-disp}, using strain, which
creates an internal gauge field acting oppositely on the two valleys
to spatially separate valley currents \cite{VPstrain-engg}, lattice
defects \cite{VPlattice-defect} and spin-orbit coupling
\cite{VPspin-orbit}.  Methods that break time-reversal include the use
of magnetic and potential barriers \cite{VPmagnetic-barrier}, or
tunnel-coupling monolayer and bilayer graphene with an in-plane
magnetic field to tune momentum \cite{VPmlg-blg}. Yet other proposals
include using adiabatic pumping \cite{VPadiabat-pump} or Floquet
methods \cite{VPfloquet,herb} to separate the valleys. Most of the
proposals need precise control of
edges/strain/substrates/superlattices, and/or depend very sensitively
on the energy of the electrons to be valley-filtered.
 \begin{figure}[t]
	\includegraphics[width=0.6\linewidth]{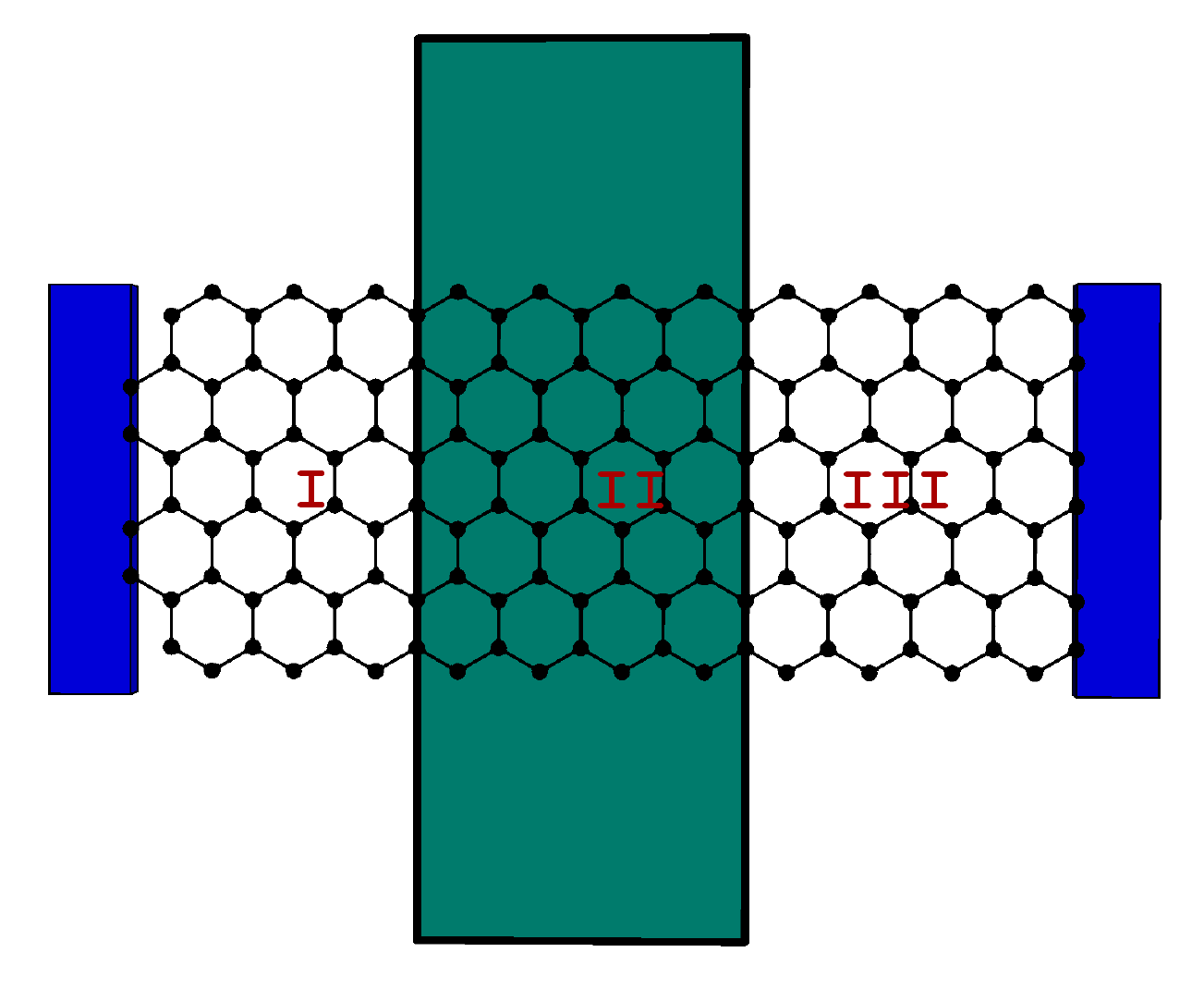}
	\caption{A schematic picture for a graphene/WSM device. The incoming current (region I) is equally populated in the two valleys and the outgoing current (region III) is valley polarized.}
	\label{fig1}
\end{figure}

\begin{figure}[t]
	\subfloat[\label{nodes1}]{%
		\includegraphics[width=0.3\linewidth]{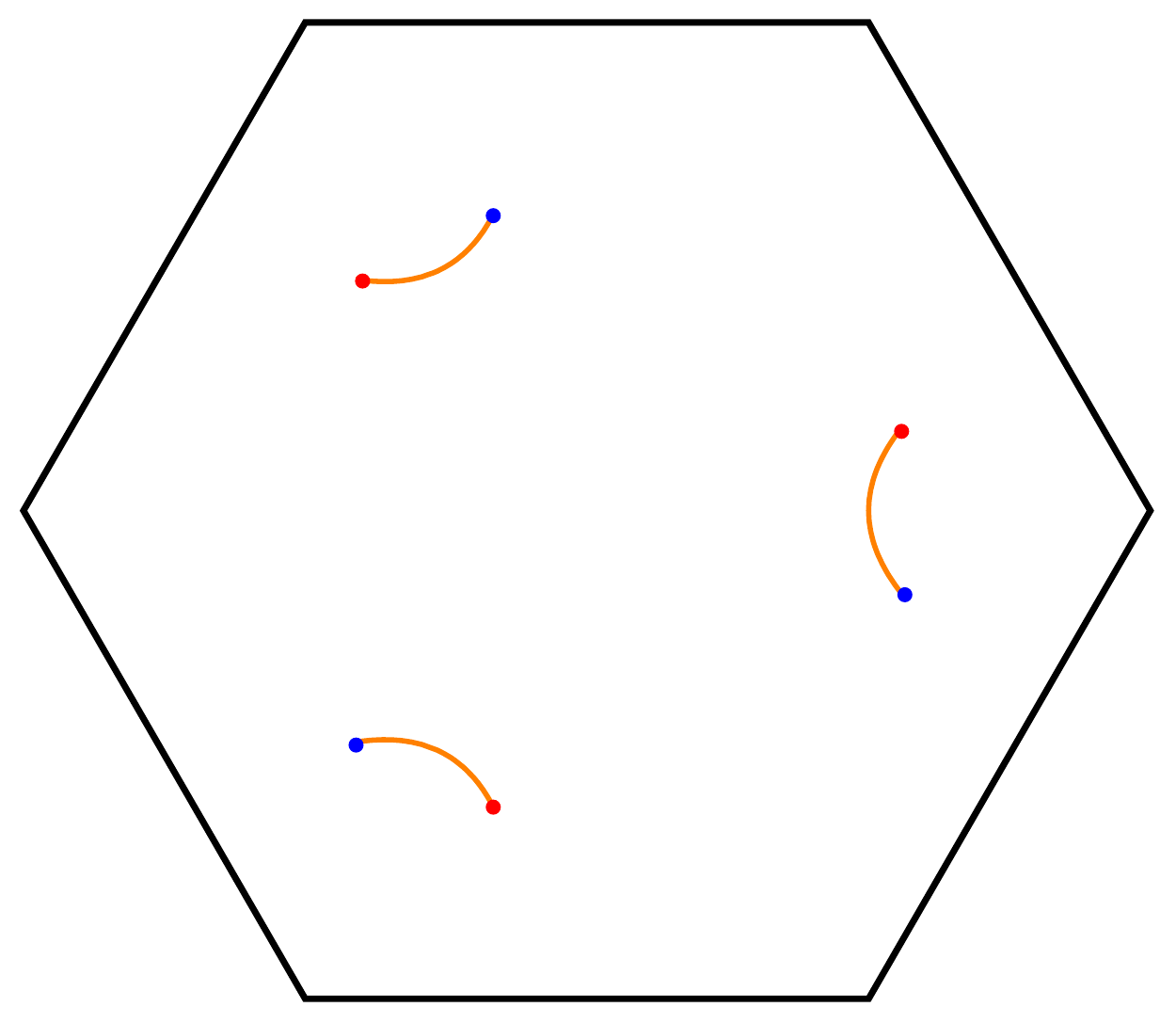}}
	\hspace{\fill}
	\subfloat[\label{nodes2} ]{%
		\includegraphics[width=0.3\linewidth]{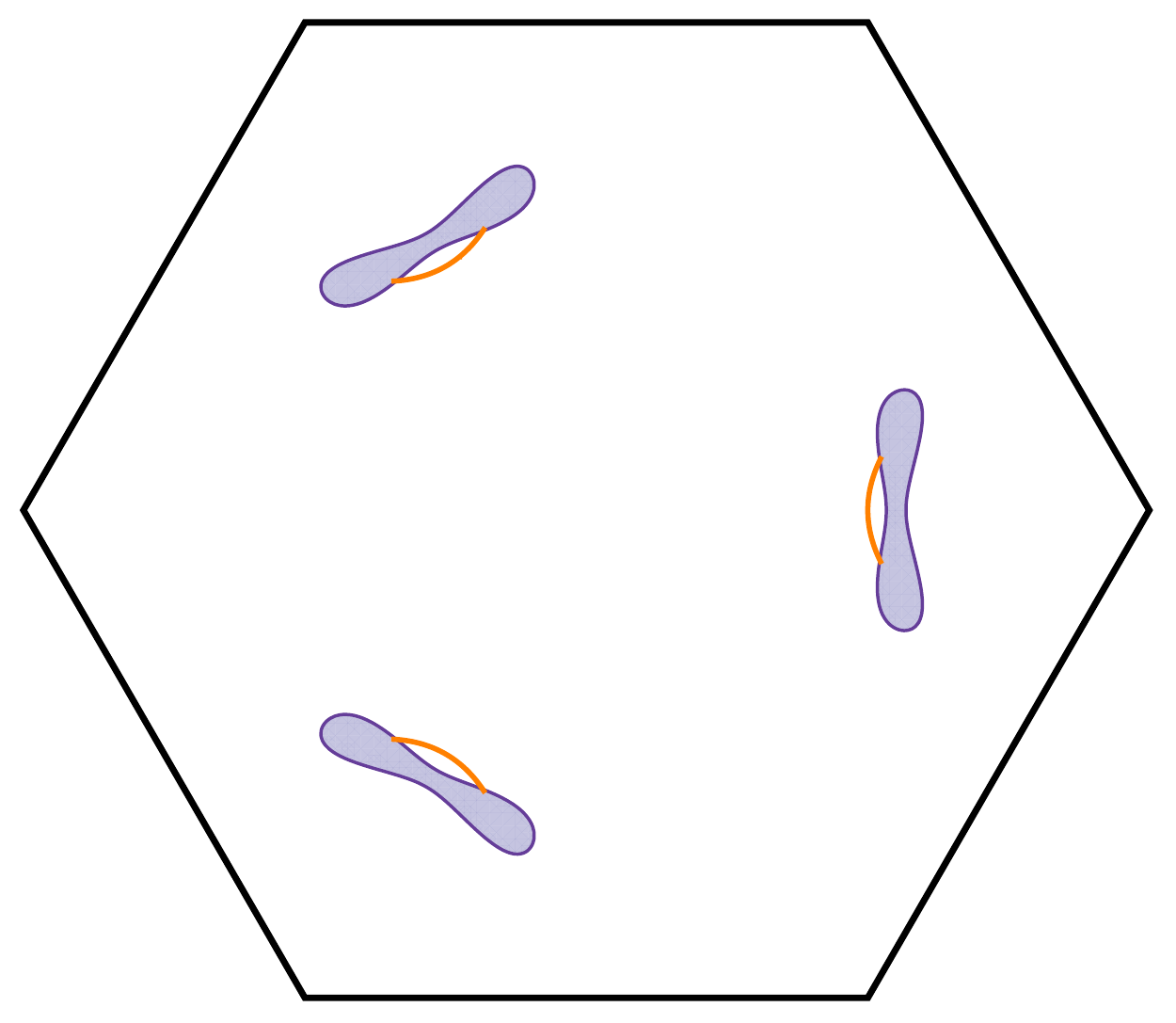}}
	\hspace{\fill}
	\subfloat[\label{nodes3}]{%
		\includegraphics[width=0.3\linewidth]{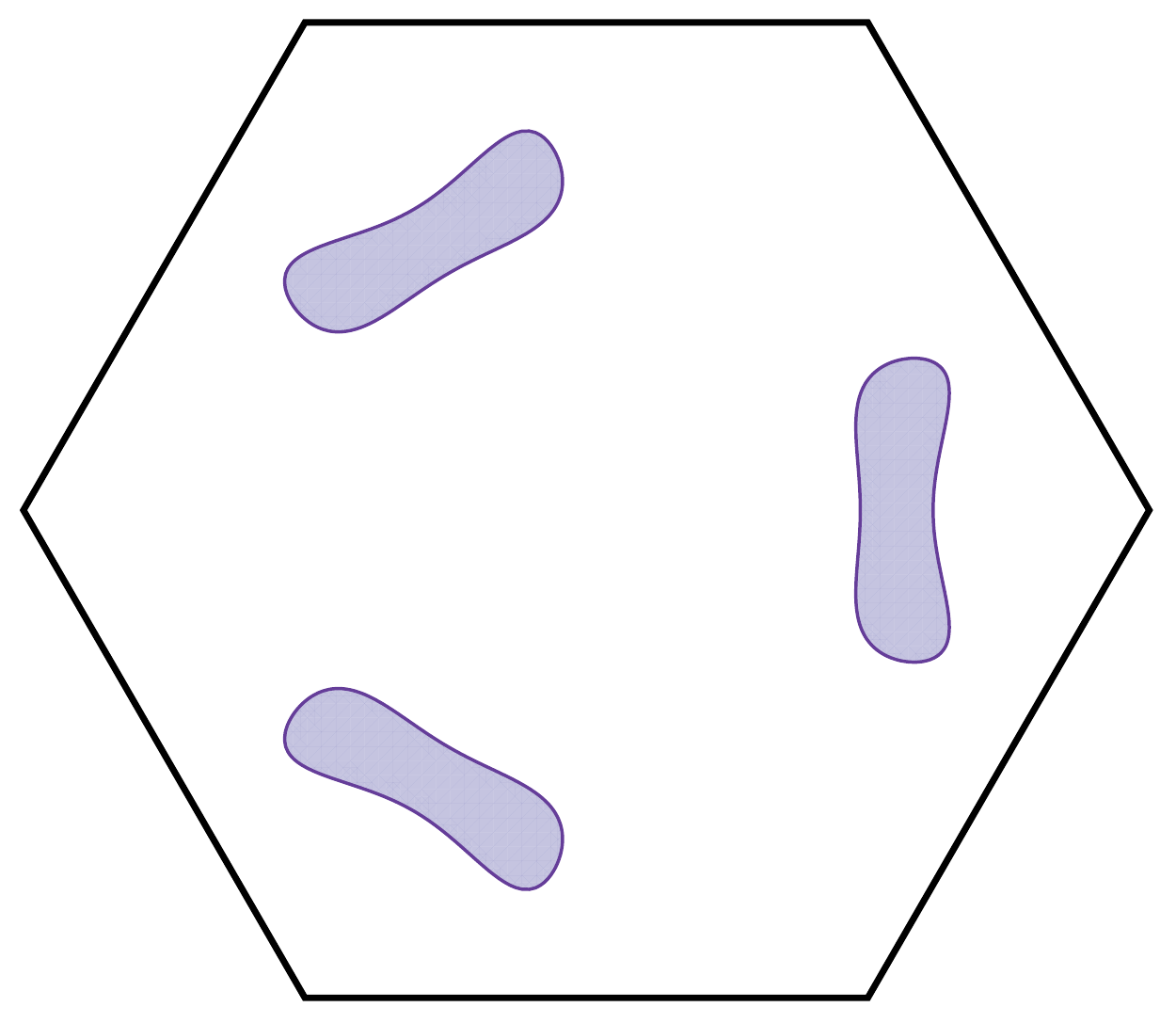}}
		\hspace{\fill}
	\subfloat[\label{gwsm}]{%
		\includegraphics[width=0.6\linewidth]{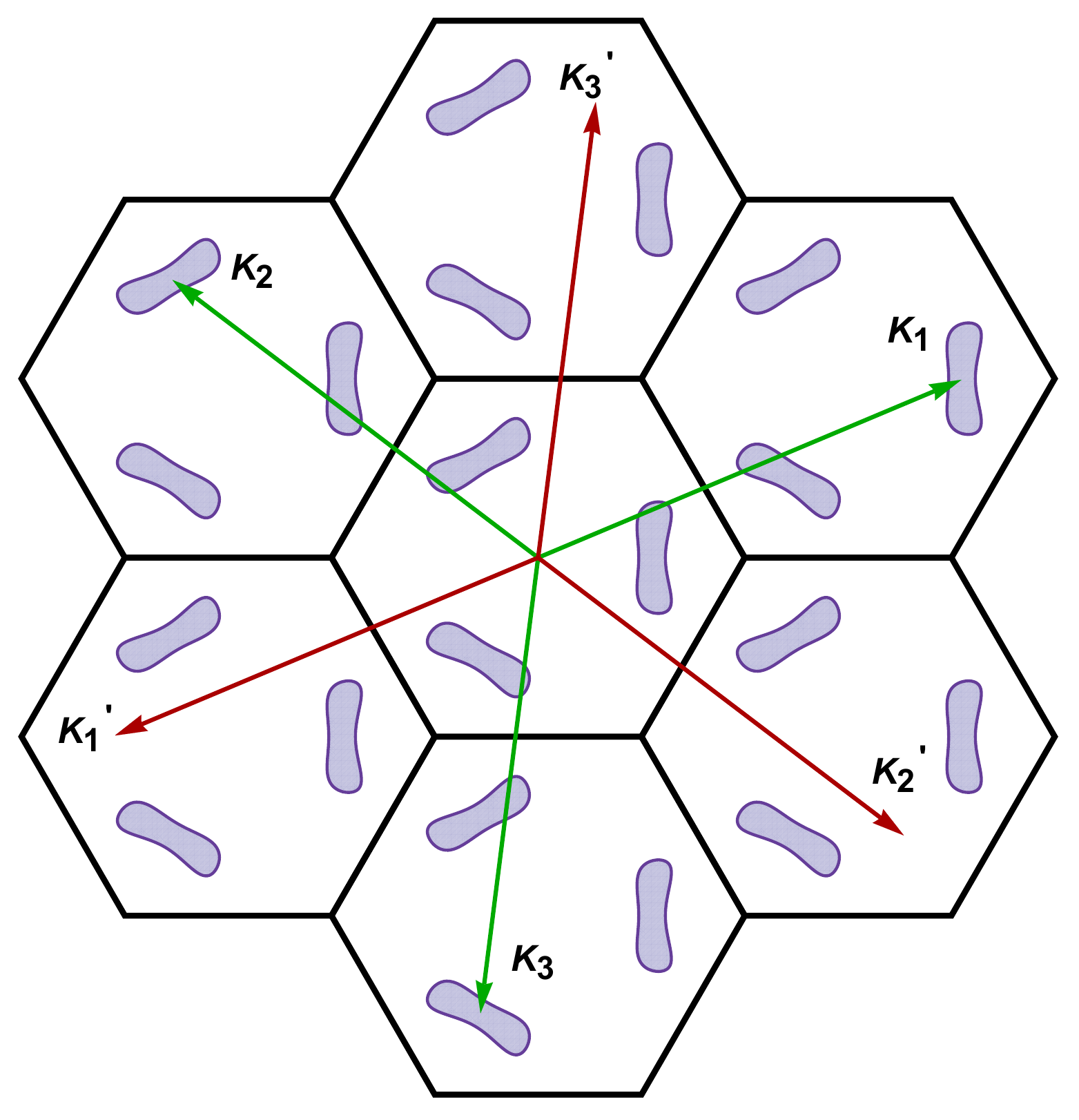}}
	\caption{\label{fig2}Evolution of the projected WSM states in
          the SBZ. (a) At the nodal energy ($E_F=0$), only the Fermi
          arcs are present in the SBZ. (b),(c) As the chemical
          potential is raised, the Fermi arcs get absorbed into the
          bulk states and eventually disappear in (c). (d) An
          incommensurate structure of graphene and the WSM shown in
          momentum space. The green vectors show the three equivalent
          graphene ${\bf K}$ points. For a range of twist angles, the
          low-energy states in the ${\bf K}$ valley of graphene lie in
          a band of the projected bulk states of the WSM. The graphene
          states near the ${\bf K}'$ points (red vectors) do not
          overlap the projected bulk states of the WSM in energy, but
          instead lie in a band gap.}
\end{figure}

In this work we show that the surface of a WSM with three-fold
symmetry, breaking both time-reversal and inversion, is a natural
substrate for robust valley-filtering current in MLG (see Fig. \ref{fig1} for the proposed device). The
minimal number of Weyl nodes is six, as seems to be the case for
Co$_3$Sn$_2$S$_2$. When the chemical potential is at the energy of the
Weyl nodes, their projections on the surface BZ (also three-fold
symmetric) are points,
%which should be
connected by zero energy surface Fermi arc (FA) states (see
Fig. \ref{fig2}\subref{nodes1}). Upon doping, the projection of the bulk states at fixed energy on to the surface BZ will be solid regions enclosing the Weyl point
projections (WPPs), as shown in Figs. \ref{fig2}\subref{nodes2}, \ref{fig2}\subref{nodes3}. We refer to these solid regions as "Fermi pockets." We
emphasize that: (i) The Fermi pockets break inversion symmetry. (ii)
Each ${\vec k}$ in the surface BZ has a continuum of bulk states of
the WSM projected on to it. The next step is to weakly tunnel-couple
the MLG to the surface of the WSM in their region of overlap, taking
care to align it so that the ${\bf K}$ Dirac point lies within a Fermi
pocket in the surface BZ of the WSM, up to a reciprocal lattice vector
of the surface BZ of the WSM, as shown in Fig. \ref{fig2}\subref{gwsm}. We emphasize that the ${\bf
  K}'$ point does not overlap a Fermi pocket. 

Consider a current injected into the graphene sheet from the left, in
region I of the device depicted in Fig. \ref{fig1}.  Before it enters
region II, where the MLG and the WSM are tunnel-coupled, the current
is divided equally between the two valleys . When it enters region II,
each electronic state in the ${\bf K}$ valley is coupled to (and lies
in the middle of) a band of bulk states in the WSM. We assume,
plausibly, that the Fermi velocity of graphene is much higher than
that of the WSM, implying that all states near the chemical potential
$\mu$ of the MLG will lie in the middle of the bulk band of the
WSM. Each state in the MLG ${\bf K}$ valley will therefore hybridize
with them and broaden, resulting in a decay of the current in the
${\bf K}$ valley into the bulk of the WSM, which is assumed to be
grounded. By contrast, although the ${\bf K}'$ valley band structure
is modified by tunnel-coupling to the WSM, there are no bulk or
surface states of the WSM at the same energy, so that the current in
this valley will suffer at most a finite diminution due to reflections
at the various interfaces of the structure. Note that the greater the
length of the tunnel-coupled region, the greater the degree of valley
polarization of the outgoing current.

A few remarks are in order about the generality and robustness of our
proposal. (i) Without additional symmetries, there is no reason for
the chemical potential in the WSM to lie at the Weyl point
energy. Thus, generically, the WSM will have Fermi pockets at the
surface. Indeed this seems to be the case for Co$_3$Sn$_2$S$_2$
\cite{Liu2019}. (ii) This implies that the alignment of the MLG on the
WSM surface can be varied over a range of angles while maintaining the
condition that ${\bf K}$ sits within a Fermi pocket, while ${\bf K}'$
does not. Thus, fine-tuning the alignment of graphene on the WSM
surface is not necessary. (iii) Scanning the chemical potential can be
achieved by doping the WSM and/or gating the MLG. Our proposal will
work over a wide range of electron energies near charge neutrality in
MLG. (iv) The details of the tunneling matrix elements between the MLG
and the surface of the WSM are irrelevant: what matters is that each
${\vec k}$ state within the ${\bf K}$ valley is coupled to the WSM
continuum. (v) Smooth disorder in the WSM or MLG will scatter
single-particle states close in momenta. Since ${\bf K}$ and ${\bf
  K}'$ are far apart, the valley-filtering will be robust against
smooth disorder.

Having established the generality and wide applicability of our
proposal, in the remainder of this paper we analyze a
specific model of such a device, illustrating the physical behaviors
described above. In order to treat arbitrary tunnel coupling strengths
via tight-binding, we construct a simple model of the three-fold
symmetric WSM, and assume that its surface is commensurate with that
of MLG. We treat only the simplest and most symmetric case in the main
text, leaving the general case to the supplemental material \cite{SM}.

\textit{WSM model.}-- Our starting point is a minimal two-band model
for a Weyl semimetal on a triangular lattice which breaks both time
reversal and inversion symmetry, but possesses three-fold rotational
symmetry. In momentum space, the Hamiltonian is given
by \begin{equation}
            \label{eq1}
	H(\vec{k},k_z)=\sum_{\mu=x,y,z}f_\mu\sigma_\mu,
\end{equation}
where
$f_x=2t\left[1-cos(k_z)+\mu_1-\sum\limits_{i=1}^{3}cos(\vec{k}.\vec{a_i})\right]$,
$f_y=2t\left[\sum\limits_{i=1}^{3}sin(\vec{k}.\vec{a_i})-\mu_2\right]$
and $f_z=2t^{'}sin(k_z)$.  Note that $\vec{k}$ here is a two
dimensional vector and $\sigma_\mu$ are the usual Pauli spin matrices,
and $t$ and $t^{'}$ represent the in-plane and out-of-plane hoppings,
respectively. The three $\vec{a}_i$ vectors are the nearest-neigbor
vectors on the triangular lattice, $\vec{a}_1=a\hat{x}$ and
$\vec{a}_{2,3}=a(\frac{-1}{2}\hat{x}\pm\frac{\sqrt{3}}{2}\hat{y})$.
The 3-fold rotational symmetry of $H$ is manifested in its energy
spectrum. The band structure possesses three pairs of Weyl nodes
related to one another by 3-fold rotations. These are found at
$k_z=0$, with $\vec{k}$ satisfying
$\mu_1-\sum\limits_{i=1}^{3}cos(\vec{k}.\vec{a_i})=0$ and
$\sum\limits_{i=1}^{3}sin(\vec{k}.\vec{a_i})-\mu_2=0$. The positions
of the Weyl nodes can be moved by varying $\mu_1$ and $\mu_2$. We
assume that the free surface of the WSM is in the $xy$ plane, which
has three-fold symmetry. The Weyl point projections (WPPs) on to the
surface Brillouin zone (SBZ) are connected by Fermi arcs. By standard
methods \cite{Murthy2020} we find the energy dispersion for the Fermi arc
states to be
\begin{equation}
\label{eq2}
	E=2\left[\sum\limits_{i=1}^{3}sin(\vec{k}.\vec{a_i})-\mu_2\right]
\end{equation}

\textit{MLG commensurate with the surface of the WSM.}-- We adopt a
model in which the MLG lattice is commensurate with that of the
surface of the WSM, and that the WSM lattice constant is smaller than that of the MLG. While these assumptions are unrealistic for real materials,
they allow us to use the full power of translation invariance to do
nonperturbative calculations in the tunnel-couplings without
fundamentally changing the character of the system.  Calculations for
incommensurate lattices are necessarily either perturbative in the
tunnel-couplings, or dependent on truncations in momentum space
\cite{Bistritzer2011,Sanjose2012}, both of which we wish to avoid. We
emphasize that our proposal for valley-filtering does not depend on
the commensuration we assume for our concrete model.

\begin{figure}[h]
	\includegraphics[width=1\linewidth]{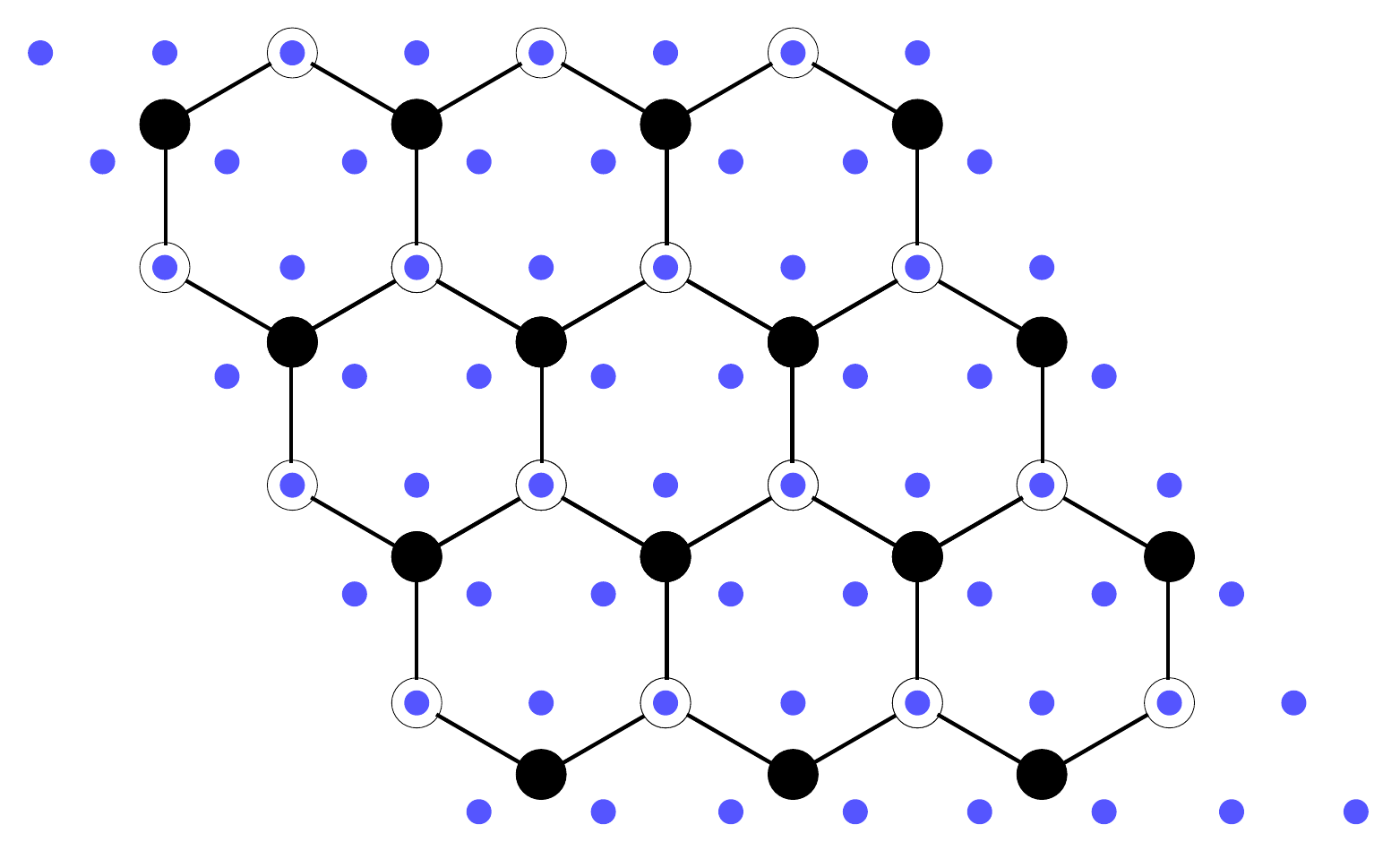}
	\caption{Graphene and the top surface of our WSM model in a
          commensurate alignment which we use in our numerical tight
          binding calculation. Electrons on the $A$ sublattice of MLG
          (empty circles) are allowed to hop only to the WSM surface
          site they overlie, while electrons on the $B$ sublattice
          (solid black) of MLG can hop to the three WSM surface sites
          surrounding the given $B$ site. }
	\label{fig3}
\end{figure}

A schematic picture of our commensurate model  is shown in
Fig. \ref{fig3}.
In order to study the electronic properties of this system, we take a
finite slab of the WSM in Eq. (\ref{eq1}) along the $z$ axis and
assume the system to be translationally invariant in the $xy$
plane. Going to real space in the $z$ direction in
Eq.  (\ref{eq1}), we obtain the WSM slab
Hamiltonian
\begin{equation}
\label{eq3}
\begin{split}
	H_{WSM} & =\sum\limits_{n=0}^{N}\sum\limits_{\vec{k}}[C_n^\dagger(\vec{k}) M(\vec{k})C_n(\vec{k})-C_{n+1}^\dagger(\vec{k}) TC_n(\vec{k})\\ 
	& -C_n^\dagger(\vec{k}) T^\dagger C_{n+1}(\vec{k})],
	\end{split}
\end{equation}
where $C_n({\vec k})$ is a two-component annihilation operator indexed by layer $n$ and
\begin{equation*}
      \begin{split}
      M(\vec{k}) & =
      2[(1+\mu_1-\sum\limits_{i}\cos(\vec{k}.\vec{a_i}))\sigma_x\\
      & +(\sum\limits_{i}\sin(\vec{k}.\vec{a_i})-\mu_2)\sigma_z].
      \end{split}
\end{equation*}
Spin has been suppressed for notational convenience. Note
that $T=\sigma_x+it^{'}\sigma_y$, $N$ is the thickness of the slab,
and that we have set the hopping in the plane our energy unit,
$t=1$. The total Hamiltonian is
\begin{equation}
       \label{eq4}
      	H=H_{WSM}+H_G+H_{t},
\end{equation}
where $H_G$ is the nearest neighbor hopping Hamiltonian of MLG. $H_t$
allows electrons in MLG to tunnnel to the top layer of the WSM in a
translation-invariant way.
\begin{equation}
         \label{eq5}
      	H_t=\sum\limits_{\vec{R}}\sum\limits_{\vec{r}}\left[C_0^\dagger(\vec{r}) V_{\alpha}(|\vec{r}-\vec{R}|)f_{\alpha}(\vec{R})+h.c\right],
\end{equation} 
where $C_0(\vec{r})$ is the two-component annihilation operator on the
top $n=0$ layer of the WSM at site $\vec{r}$, and
$f_{\alpha}(\vec{R})$ is an annihilation operator on the sublattice
$\alpha=A,B$ at site $\vec{R}$ in graphene.\\

To ensure our requirement that the neighborhood of the ${\bf K}$ point
of MLG lies within a bulk band of energies of the WSM we assume that
the ${\bf K}$ point of MLG lies on a Fermi arc.  From Eq. (\ref{eq2}),
this is achieved when $\mu_2=-\sqrt{3}/2$. The ${\bf K}'$ point will
then reside in the gap of the WSM. We then diagonalize Eq. (\ref{eq4})
for this value of $\mu_2$ to get the band structure of the system. We
restrict ourselves to nearest-neighbor hopping only in $H_t$. This
operationally means that electrons on the $A$ sublattice of MLG hop
only to the WSM surface site at the same $xy$ coordinates with a
spin-independent amplitude $\kappa$, while electrons on the $B$
sublattice of MLG can hop to the three sites of the WSM surface
surrounding it with spin-independent amplitude $\kappa'$. As we show
in the supplemental material \cite{SM}, moving the Fermi arcs or
making the hopping more generic does not make any qualitative
difference to our results. \\
       
\begin{figure}[t]
\captionsetup[subfigure]{labelformat=empty}
	\subfloat[\label{kvalley}]{%
		\includegraphics[width=0.475\linewidth]{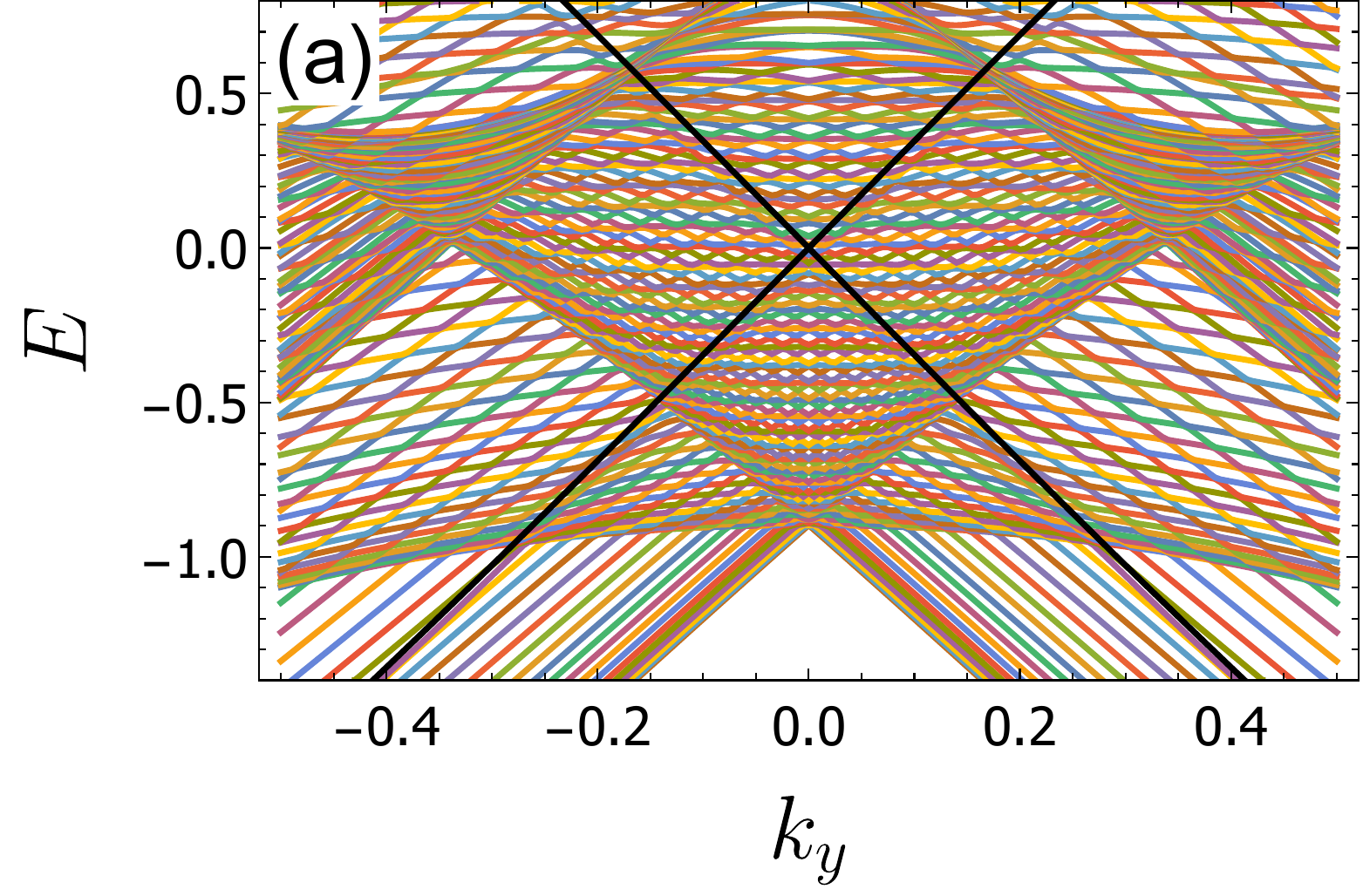}}
	\hspace{1em}
	\subfloat[\label{kpvalley}]{%
		\includegraphics[width=0.475\linewidth]{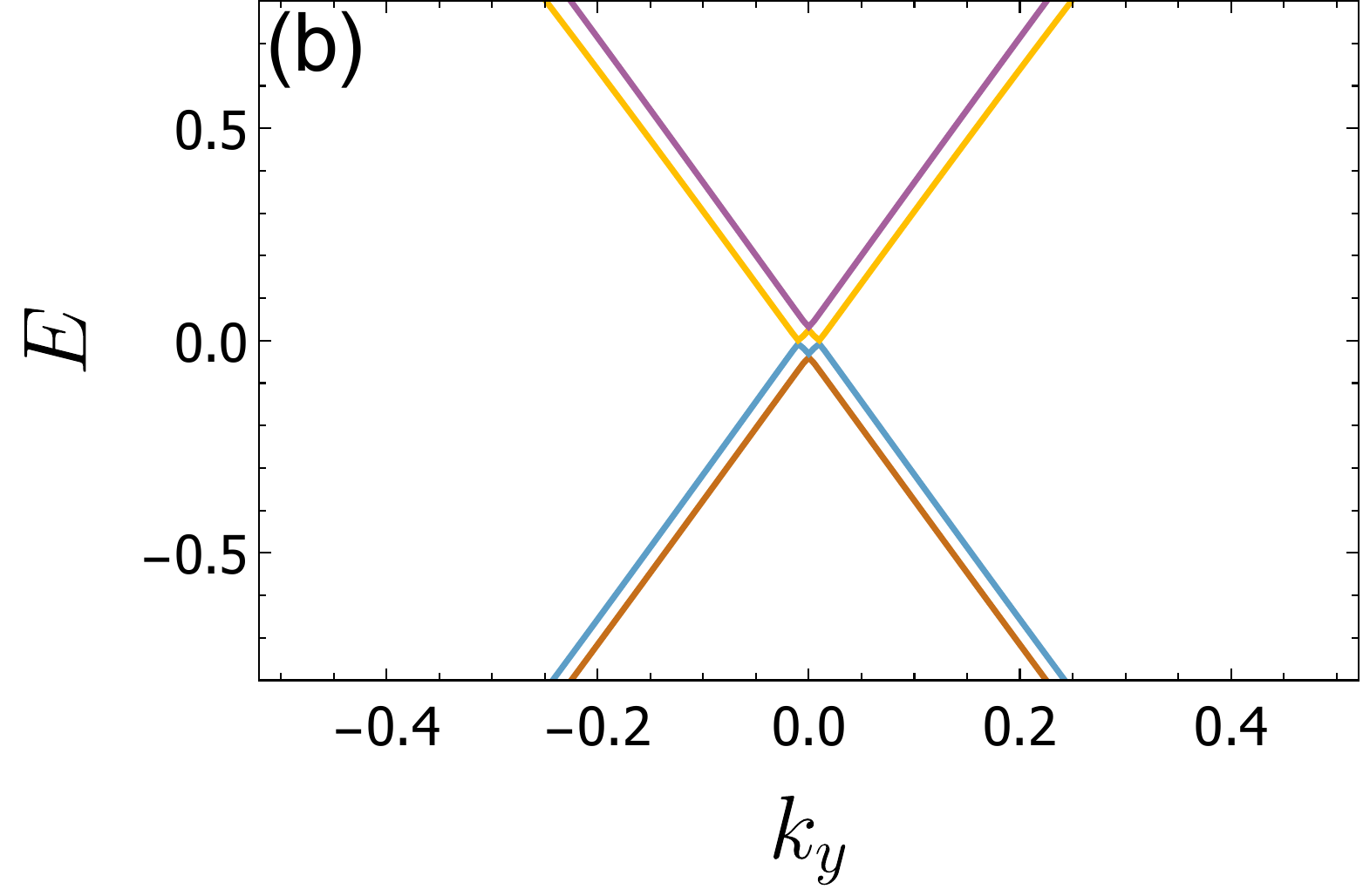}}
	\hspace{1em}	
	\subfloat[\label{close}]{%
		\includegraphics[width=0.475\linewidth]{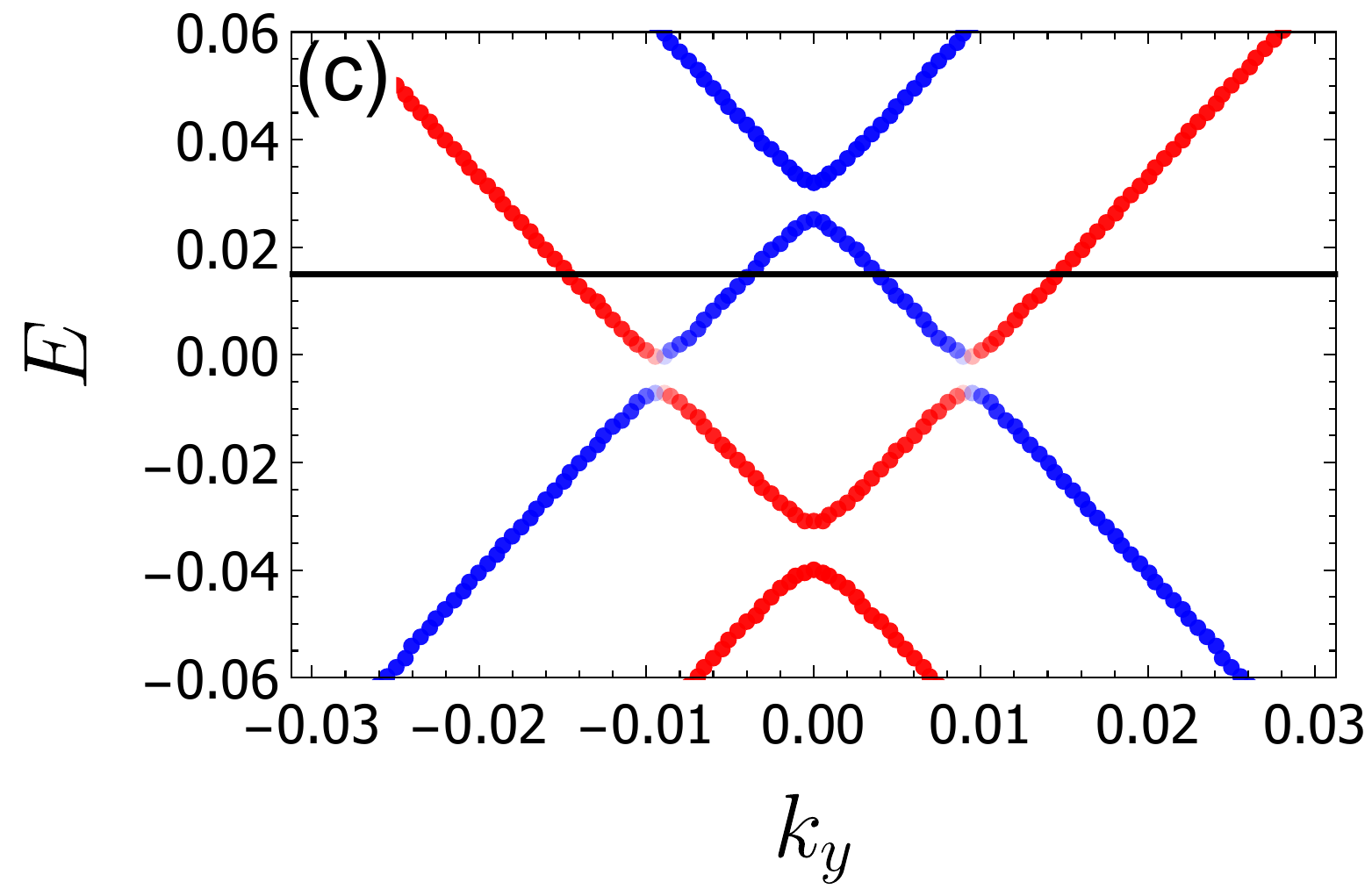}}
	\hspace{1em}	
	\subfloat[\label{cond}]{%
		\includegraphics[width=0.475\linewidth]{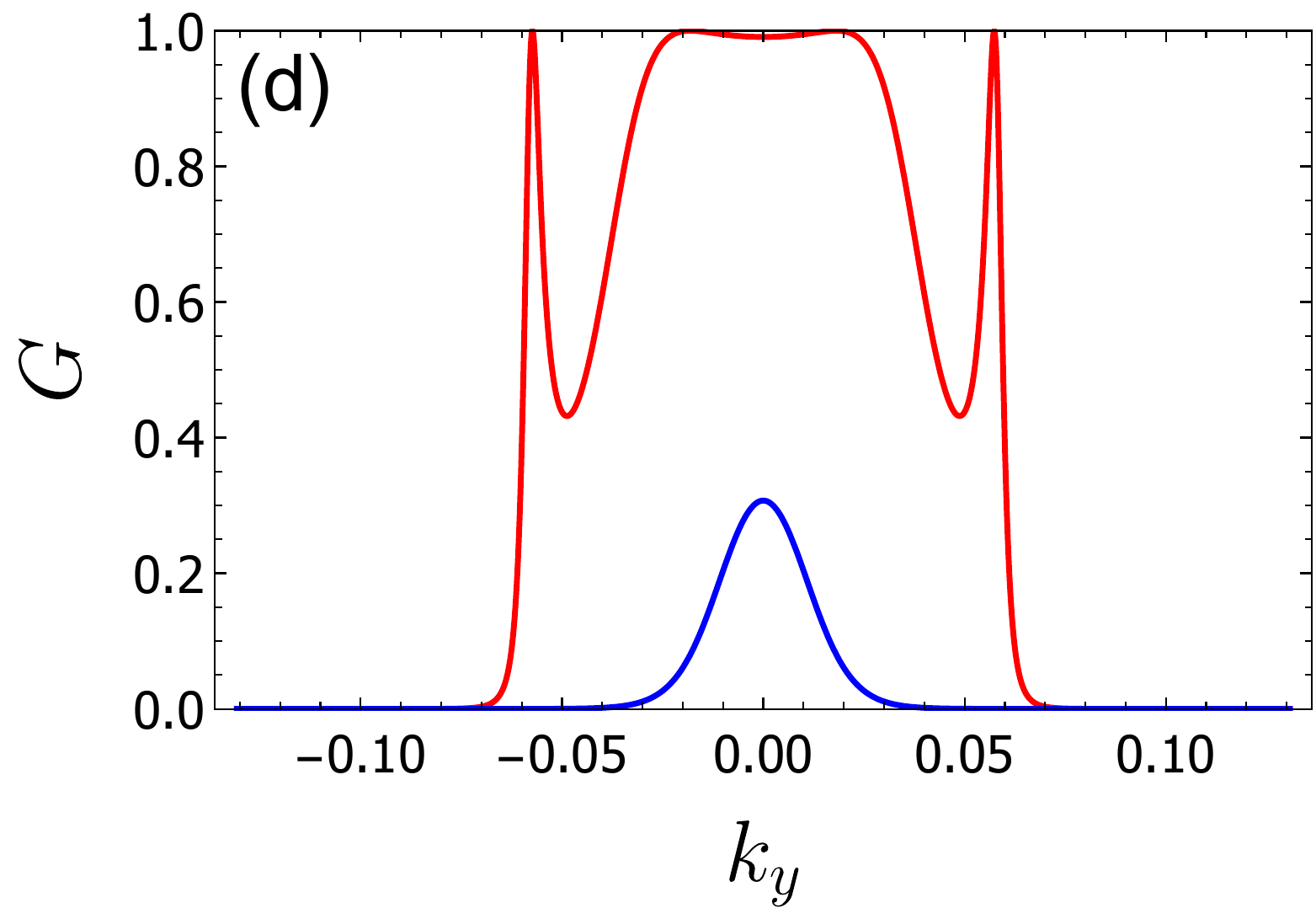}}\\
	\caption{\label{fig4}Results of the tight binding
          calculation. (a),(b) The graphene/WSM spectrum is shown near
          the ${\bf K}$ and the ${\bf K}'$ valleys, respectively. The
          Dirac cone in the ${\bf K}$ valley (shown in black) is
          immersed in the bulk states of the WSM while the Dirac
          cone near the ${\bf K}'$ valley is slightly perturbed. (c)
          An expanded view zooming in on the ${\bf K}'$ valley shows an
          inverted band structure where the colors code the average of
          $S_z$ operator with red being spin up and blue spin
          down. (d) The conductance of the graphene/WSM device based
          on a simplified model for the states near the ${\bf K}'$
          valley is shown in units of $e^2/h$. The conductance shows
          partial spin polarization as red denotes spin up states
          where blue denotes spin down states.}
\end{figure} 

Our tight-binding calculations presented in
Fig. \ref{fig4}\subref{kvalley},\subref{kpvalley} are consistent with
expectations from the generic incommensurate case discussed
earlier. We see that the Dirac cone for the ${\bf K}$ valley is
immersed in the continuum of the bulk states while the ${\bf K}'$
valley is isolated in the gap.  A close examination of
Fig. \ref{fig4}\subref{kpvalley} reveals an induced spin-orbit (SO)
coupling in the states of graphene's ${\bf K}'$ valley.

The tight-binding results show that, at the ${\bf K}'$ valley, the
bands split into (almost purely) $\uparrow$ and $\downarrow$ bands
with a band inversion near the ${\bf K}'$ point
(Fig. \ref{fig4}\subref{close}).  This can lead to interesting
consequences for the transport: the different sizes of the Fermi
surfaces for the two spin species (at the Fermi level marked black in
Fig. \ref{fig4}\subref{close}) will lead to different transmission
probabilities, and consequently a (partial) spin polarization of the
fully valley polarized current.
     
A simple model that reproduces the tight-binding results is the
following effective Hamiltonian:
     \begin{equation}
        \label{eq6}
     	H_{eff}(\vec{k})=-k_{x}\tau_{x}+k_{y}\tau_{y}+\lambda_{1}\tau_z-\lambda_{2}\sigma_{z},
     \end{equation} 
 where $\tau$ and $\sigma$ denote Pauli matrices corresponding to the
 sublattice and spin degrees of freedom, respectively, while
 $\lambda_{1}$ and $\lambda_{2}$ are the sublattice- and
 time-reversal-breaking parameters induced by integrating out the
 gapped WSM states. 
    
 Fig. \ref{fig4}\subref{cond} shows the Landauer conductance
 ($G=\frac{e^2}{h}T$, where $T$ is the transmission probability; see
 the Supplemental Material \cite{SM} for details) of our proposed
 device (Fig. \ref{fig1}) in the two spin channels. We have assumed
 that there is translation invariance in the $y$-direction,
 perpendicular to the current flow. Given our assumptions, the
 conductance is due to the ${\bf K}'$ valley only. %Details of the
 %calculations are given in the Supplemental Material \cite{SM}. 
 Since the
 coupling to the WSM breaks both the sublattice and time-reversal
 symmetries of MLG, it is natural that the conductances in the two
 spin channels are unequal.

 %However, while we do expect a generic nonzero
 %spin-polarization of the transmitted current, its precise value will
 %depend on the effective parameters $\lambda_{1,2}$.
 %and is thus not expected to be robust.  
 %Detailed examination of this dependence is left for future investigation.

\textit{Conclusions.}--We have shown that overlaying graphene on a
3-fold symmetric surface of a WSM breaking both time-reversal and
inversion, with an alignment which places the ${\bf K}$ point of MLG
in a Fermi pocket of the surface BZ of the WSM (modulo reciprocal
lattice vectors of the WSM), will lead to a robust valley filter for
graphene.

The physics leading to this may be stated concisely: states near the
${\bf K}$ valley of MLG lie within a band of bulk states of the WSM,
projected to the surface BZ, hybridizing with the bulk states and
``dissolving'' into them. Current-carrying electrons in the ${\bf K}$
valley will scatter into bulk states of the (grounded) WSM, carrying
them away from the MLG layer. States near the ${\bf K}'$ valley, on
the other hand, lie in a bandgap of the WSM, and will remain localized
in the MLG, though their transport will be modified by the sublattice
and time-reversal breaking induced by the WSM. Thus, for a
sufficiently long interface (along the current direction), only the
current in the ${\bf K}'$ valley survives. This current is expected to
have a nonzero spin polarization, whose precise value depends on the
details of the interface coupling.

Our proposal does not require precise alignment between graphene and
the surface of the WSM, precise control of the tunneling at the
interface or the chemical potential of the current-carrying
electrons. Smooth disorder will not degrade the valley-filtering. Upon
doping the WSM appropriately, our proposal will work for
Bernal-stacked/twisted bilayer graphene \cite{GrapheneReview} and
transition-metal-dichalcogenides \cite{TMD-review} as well.

Diverse applications in addition to valley-filtering can also be considered. With
small changes, the WSM could be used as a contact that is electrically
connected only to one valley, which could be used to probe equilibrium
correlated states in the quantum Hall regime of MLG. It would also be
interesting to ask how the correlated states in magic-angle twisted
bilayer graphene \cite{TBLG-expt} respond when the states near one
valley dissolve into the WSM bulk. We hope to address these and other
questions in the near future.

  We acknowledge  support from the National Science Foundation via
grant nos. DMR-2026947 (A.M.K and R.K.K.),  ECCS-1936406 and DMR-1914451 (H.A.F.), as well as the support of  the 
Research  Corporation  for  Science  Advancement through a Cottrell SEED
Award (H.A.F); the US-Israel Binational Science Foundation through awards No.
2016130 (H.A.F, G.M. and E.S.) and 2018726 (H.A.F and E.S.); and the Israel Science Foundation (ISF)
Grant No. 993/19 (E.S.).


\begin{thebibliography}{99}
 
 \bibitem{Murakami2007} S. Murakami, \href{https://iopscience.iop.org/article/10.1088/1367-2630/9/9/356/meta}{New J. Phys. {\bf 9}, 356 (2007).} 
 
 \bibitem{Wan2011} X.~Wan, A.~M.~Turner, A.~Vishwanath and S.~Y.~Savrasov,
\href{https://journals.aps.org/prb/abstract/10.1103/PhysRevB.83.205101}{Phys. Rev. B {\bf 83}, 205101 (2011)};  P.~Hosur, S.~A.~Parameswaran and A. Vishwanath, \href{https://journals.aps.org/prl/abstract/10.1103/PhysRevLett.108.046602}{\prl\ {\bf 108}, 046602 (2012).}

 \bibitem{Yang2011} K. Y.  Yang, Y. M. Lu and Y. Ran, \href{https://journals.aps.org/prb/abstract/10.1103/PhysRevB.84.075129}{Phys. Rev. B {\bf 84}, 075129 (2011).}

 \bibitem{Burkov2011} A.~A.~Burkov and L.~Balents, \href{https://journals.aps.org/prl/abstract/10.1103/PhysRevLett.107.127205}{\prl\ {\bf 107}, 127205 (2011).}

 \bibitem{Xu2011} G. Xu, H. Weng, Z. Wang, X. Dai, and Z. Fang, \href{https://journals.aps.org/prl/abstract/10.1103/PhysRevLett.107.186806}{\prl\
  {\bf 107}, 186806 (2011).}

 \bibitem{Lv2015} B. Q. Lv, H. M. weng, B. B. Fu, X. P. Wang, H. Miao,
  J. Ma, P. Richard, X. C. Huang, L. X. Zhao, G. F. Chen {\it et al},
  \href{https://journals.aps.org/prx/abstract/10.1103/PhysRevX.5.031013}{Phys. Rev. X  {\bf 5}, 031013 (2015).}

 \bibitem{Xu2015a} S.-Y. Xu, I. Beloploski, N. Alidoust, M. Neupane,
  G. Bian, C. Zhang, R. Sankar, G. Chang, Z, Yuan, C.-C. Lee {\it et
    al}, \href{https://science.sciencemag.org/content/349/6248/613.abstract}{Science {\bf 349}, 613 (2015).}

 \bibitem{Xu2015b} S.-Y. Xu, N. Alidoust, I. Blopolski, Z. Yuan,
  G. Bian, T.-R. Chang, H. Zheng, V. N. Strocov, D. S. Sanchez,
  G. Chang {\it et al}, \href{https://www.nature.com/articles/nphys3437}{Nat. Phys. {\bf 11}, 748 (2015).}

 \bibitem{Inoue2016} H. Inoue, A. Gyenis, Z. Wang, J. Li, S. W. Oh,
  S. Jiang, N. Ni, B. A. Bernevig, and A. Yazdani, \href{https://science.sciencemag.org/content/351/6278/1184.abstract}{Science {\bf 351},
  1184 (2016).}

 \bibitem{SilvaNeto2019} E. H. Silva Neto, \href{https://science.sciencemag.org/content/365/6459/1248/tab-article-info}{Science {\bf 365}, 1248
  (2019).}

 \bibitem{Belopolski2019} I. Belopolski, K. Manna, D. S. Sanchez,
  G. Chang, B. Ernst, J. Yin, S. S. Zhang, T. Cochran, N. Shumiya,
  H. Zheng {\it et al}, \href{https://science.sciencemag.org/content/365/6459/1278/tab-article-info}{Science {\bf 365}, 1278 (2019).}

 \bibitem{Liu2019} D. F. Liu, A. J. Liang, E. K. Liu, Q. N. Xu,
  Y. W. Li, C. Chen, D. Pei, W. J. Shi, S. K. Mo, P. Dudin {\it et
    al}, \href{https://science.sciencemag.org/content/365/6459/1282}{Science {\bf 365}, 1282 (2019).}

 \bibitem{Morali2019} N. Morali, R. Batabyal, P. K. Nag, E. Liu, Q. Xu,
  Y. Sun, B. Yan, C. Felser, N. Avraham, and H. Beidenkopf, \href{https://science.sciencemag.org/content/365/6459/1286.editor-summary}{Science
  {\bf 365}, 1286 (2019).}

 \bibitem{GrapheneReview} A. H. Castro Neto, F. Guinea, N. M. R. Peres, K. S. Novoselov, and A. K. Geim, \href{https://journals.aps.org/rmp/abstract/10.1103/RevModPhys.81.109}{Rev. Mod. Phys. {\bf 81}, 109 (2009).}

 \bibitem{ValleytronicsReview} J. R. Schaibley, H. Yu, G. Clark,
  P. Rivera, J. S. Ross, K. L. Seyler, W. Yao, and X. Xu, \href{https://www.nature.com/articles/natrevmats201655}{Nature
  Reviews Materials {\bf 1}, 16055 (2016).}

 \bibitem{Xiao2007} D. Xiao, W. Yao, and Q. Niu, \href{https://journals.aps.org/prl/abstract/10.1103/PhysRevLett.99.236809}{\prl\ \bf{99}, 236809 (2007).} 
  
 \bibitem{Rycerz2007} A. Rycerz, J. Tworzydlo, and C. W. J. Beenakker,
  \href{https://www.nature.com/articles/nphys547}{Nature Physics {\bf 3}, 172 (2007).}

 \bibitem{trigonal-disp} J. L. Garcia-Pomar, A. Cortijo, and
  M. Nieto-Vesperinas, \href{https://journals.aps.org/prl/abstract/10.1103/PhysRevLett.100.236801}{\prl\ {\bf 100}, 236801 (2008)}; Y. S. Ang,
  S. A. Yang, C. Zhang, Z. Ma, and L. K. Ang, \href{https://journals.aps.org/prb/abstract/10.1103/PhysRevB.96.245410}{Phys. Rev. B. {\bf 96}, 245410
  (2017).}

 \bibitem{VPstrain-engg}  Z.-P, Niu, \href{https://aip.scitation.org/doi/10.1063/1.4720386}{Jour. Appl. Phys. {\bf 111}, 103712 (2012)}; 
S. P. Milovanovic and F. M. Peeters, \href{https://aip.scitation.org/doi/abs/10.1063/1.4967977}{Appl. Phys. Lett. {\bf 109}, 203108 (2016)}; 
M. Settnes, S. R. Power, M. Brandbyge, and A.-P. Jauho, \href{https://journals.aps.org/prl/abstract/10.1103/PhysRevLett.117.276801}{\prl\ {\bf 117}, 276801 (2016)}; 
T. Stegmann and N. Szpak, \href{https://iopscience.iop.org/article/10.1088/2053-1583/aaea8d}{2D Materials {\bf 6}, 015024 (2019).}

 \bibitem{VPlattice-defect} D. Gunlycke and C. T. White, \href{https://journals.aps.org/prl/abstract/10.1103/PhysRevLett.106.136806}{\prl\ {\bf 106}, 136806 (2011)}; 
  L. H. Ingaramo and L. E. F. Foa Torres, \href{https://iopscience.iop.org/article/10.1088/0953-8984/28/48/485302}{J. Phys. Condens. Matter {\bf 28}, 485302 (2016).}

 \bibitem{VPspin-orbit} M. M. Grujic, M. Z. Tadic, and F. M. Peeters,
  \href{https://journals.aps.org/prl/abstract/10.1103/PhysRevLett.113.046601}{\prl\ {\bf 113}, 046601 (2014).}

 \bibitem{VPmagnetic-barrier} D. Moldovan, M. Ramezani Masir,
  L. Covaci, and F. M. Peeters, \href{https://journals.aps.org/prb/abstract/10.1103/PhysRevB.86.115431}{\prb\ {\bf 86}, 115431 (2012)}; F. Zhai,
  Y. Ma, and Y.-T. Zhang, \href{https://iopscience.iop.org/article/10.1088/0953-8984/23/38/385302}{JPCM {\bf 23}, 385302 (2011)}; F. Zhai,
  \href{https://pubs.rsc.org/en/content/articlelanding/2012/nr/c2nr31701j#!divAbstract}{Nanoscale {\bf 4}, 6527 (2012)}; F. Zhai and K. Chang, \href{https://journals.aps.org/prb/abstract/10.1103/PhysRevB.85.155415}{\prb\ {\bf 85},
  155415 (2012)}; Y. Wang, \href{https://aip.scitation.org/doi/10.1063/1.4818607}{J. App. Phys. {\bf 114}, 073709 (2013)};
  W.-T. Lu, \href{https://journals.aps.org/prb/abstract/10.1103/PhysRevB.94.085403}{\prb\ {\bf 94}, 085403 (2016)}; M. Settnes, J. H. Garcia, and
  S. Roche, \href{https://iopscience.iop.org/article/10.1088/2053-1583/aa7cbd}{2D Materials {\bf 4}, 031006 (2017)}; J. Wang, M. Long,
  W.-S. Zhao, G. Wang, and K. S. Chan, \href{https://iopscience.iop.org/article/10.1088/0953-8984/28/28/285302}{J. Phys. Condens. Matter {\bf
    28}, 285302(2016)}; Q.-P. Wu, Z.-F. Liu, A.-X. Chen, X.-B. Xiao,
  and Z.-M. Liu, Scientific Reports {\bf 6}, 21950 (2016); M. M. Asmar
  and S. E. Ulloa, \href{https://journals.aps.org/prb/abstract/10.1103/PhysRevB.97.241104}{\prb\ {\bf 96}, 201407(R) (2017)}; A. R. S. Lins and
  J. R. F. Lima, \href{https://www.sciencedirect.com/science/article/abs/pii/S0008622320300312}{Carbon {\bf 160}, 353 (2020)}; Q.-P. Wu, L.-L. Chang,
  Y.-Z. Li, X.-B. Xiao, Z.-F. Liu, \href{https://www.sciencedirect.com/science/article/abs/pii/S1386947719303327}{Physica E {\bf 118}, 113864 (2020).}

 \bibitem{VPmlg-blg} L. Pratley and U. Zulicke, \href{https://aip.scitation.org/doi/abs/10.1063/1.4866591}{App. Phys. Lett. {\bf
  104}, 082401 (2014).}

 \bibitem{VPadiabat-pump} Y. Jiang, T. Low, K. Chang, M. I. Katsnelson,
  and F. Guinea, \href{https://journals.aps.org/prl/abstract/10.1103/PhysRevLett.110.046601}{\prl\ {\bf 110} (2013).}

 \bibitem{VPfloquet}   F. Qi and G. Jin, \href{https://aip.scitation.org/doi/10.1063/1.4874676}{J. App. Phys. {\bf 115}, 173701 (2014).}
 
 \bibitem{herb} A. Kundu, , H.A. Fertig, and B. Seradjeh, \href{https://aip.scitation.org/doi/10.1063/1.4874676}{\prl\ {\bf 116} 016802 (2016).}
 
 %\bibitem{Mah00} See, for example, G. D. Mahan, %\href{https://www.springer.com/gp/book/9780306463389}{Many Particle Physics, Third Edition (Plenum, New York, 2000).}

\bibitem{Murthy2020} G.~Murthy, H. A.~Fertig, and E.~Shimshoni, \href{https://journals.aps.org/prresearch/abstract/10.1103/PhysRevResearch.2.013367}{Phys. Rev. Research {\bf 2}, 013367(2020).}

\bibitem{Bistritzer2011} R. Bistritzer and A. H. MacDonald,
  \href{https://www.pnas.org/content/108/30/12233}{Proc. Nat. Acad. Sci. USA {\bf 108}, 12233 (2011).}

\bibitem{Sanjose2012} P. San-Jose, J. Gonzalez, and F. Guinea, \href{https://journals.aps.org/prl/abstract/10.1103/PhysRevLett.108.216802}{\prl
  {\bf 108}, 216802 (2012).}

\bibitem{SM} Supplementary Material.

\bibitem{TMD-review} For a review, see, S. Manzeli, D. Ovchinnikov,
  D. Pasquier, O. V. Yazyev, and A. Kis, \href{https://www.nature.com/articles/natrevmats201733}{Nat. Rev. Mater. {\bf
    2}, 17033 (2017).}
  
\bibitem{TBLG-expt} Y. Cao, V. Fatemi, S. Fang, K. Watanabe,
  T. Taniguchi, E. Kaxiras, R. C. Ashoori, and P. Jarillo-Herrero,
  \href{https://www.nature.com/articles/nature26160}{Nature {\bf 556}, 43 (2018)}; Y. Cao, V. Fatemi, S. Fang,
  S. L. Tomarken, J. Y. Luo, J. D. Sanchez-Yamagishi, K. Watanabe,
  T. Taniguchi, E. Kaxiras, R. C. Ashoori, and P. Jarillo-Herrero,
  \href{https://www.nature.com/articles/nature26154}{Nature {\bf 556}, 80 (2018).}
  
 
\end{thebibliography}
\end{document}

\title{Weyl Semimetal Path to Valley Filtering in Graphene: Supplemental Material}
\author{Ahmed M. Khalifa}
\affiliation{Department of Physics and Astronomy, University of Kentucky, Lexington, KY 40506-0055}
\author{Ribhu K. Kaul}
\affiliation{Department of Physics and Astronomy, University of Kentucky, Lexington, KY 40506-0055}
\author{Efrat Shimshoni}
 \affiliation{Department of Physics, Bar-Ilan University,
  Ramat-Gan 52900, Israel}
\author{H.A. Fertig}
  \affiliation{Department
  of Physics, Indiana University, Bloomington, IN 47405}
\author{Ganpathy Murthy}
\affiliation{Department of Physics and Astronomy, University of Kentucky, Lexington, KY 40506-0055}
	\maketitle

In this set of supplemental materials we present calculations
substantiating statements made in the main text, and also present
evidence of the robustness of valley-filtering to various
perturbations to the highly symmetric model considered in the main
text. That model retained the three-fold symmetry of the Hamiltonian
in the hopping between the monolayer graphene (MLG) and the top layer
of the Weyl semimetal (WSM). Further, that hopping was assumed to be
spin-independent. In Section \ref{SM:sec1} we consider the most
general hopping, which breaks the three-fold rotation symmetry and
allows for spin flips during hopping. We present its effects on the
spectrum in the ${\bf K}'$ valley. In Section \ref{SM:sec2} we present
numerical evidence for our main claim, that MLG states in the ${\bf
  K}$ valley ``dissolve'' into the bulk states of the WSM
\cite{Mah00}. In Section \ref{SM:sec3} we show the details of the
calculation of the Landauer conductance of the ${\bf K}'$
valley. Finally, in Section \ref{SM:sec4} we treat a highly fine-tuned
case, which is nevertheless quite interesting. This is the case when
the energies of the MLG Dirac point and the WSM Weyl points coincide,
and also lie at the chemical potential. A nontrivial reconstruction of
the Fermi arcs takes place.
\section{Generic hopping between Graphene and the WSM surface}
\label{SM:sec1}
\begin{figure}[h]
	\subfloat[\label{str}]{%
		\includegraphics[width=0.3\linewidth]{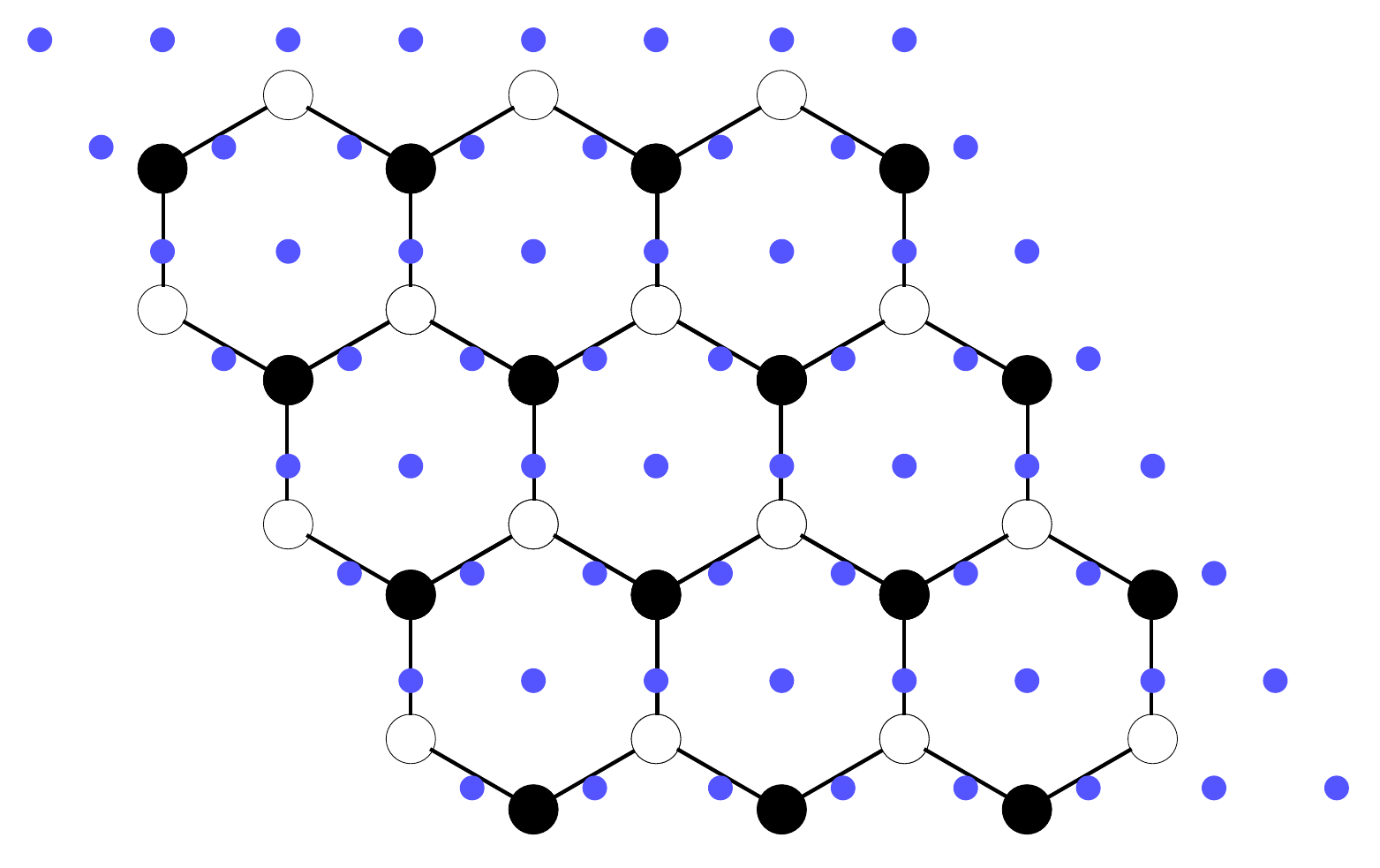}}
	\hspace{1em}
	\subfloat[\label{c3break1}]{%
		\includegraphics[width=0.3\linewidth]{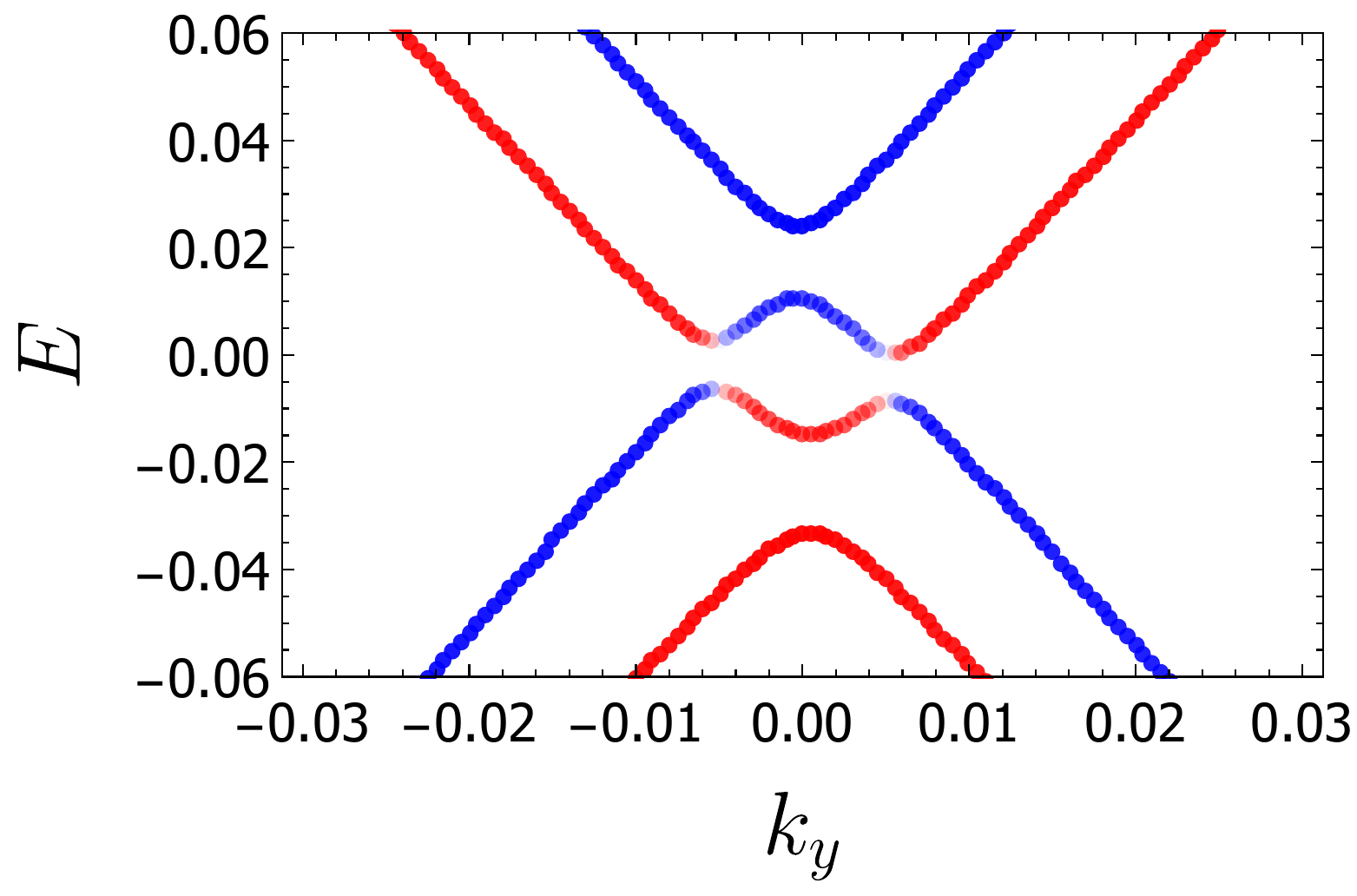}}
	\hspace{1em}
	\subfloat[\label{c3break2}]{%
		\includegraphics[width=0.3\linewidth]{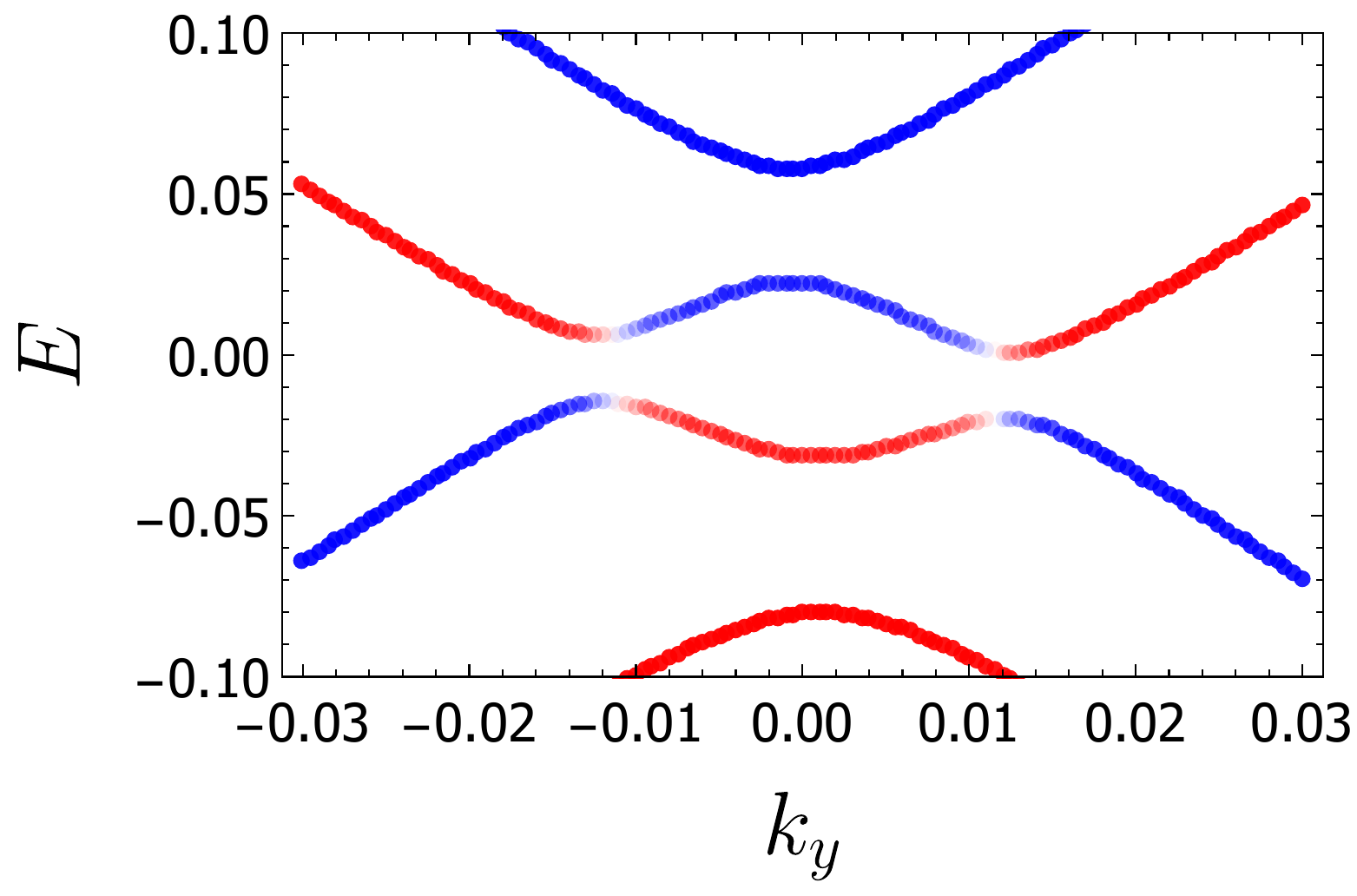}}
	\hspace{1em}
	\subfloat[\label{c3break3}]{%
		\includegraphics[width=0.3\linewidth]{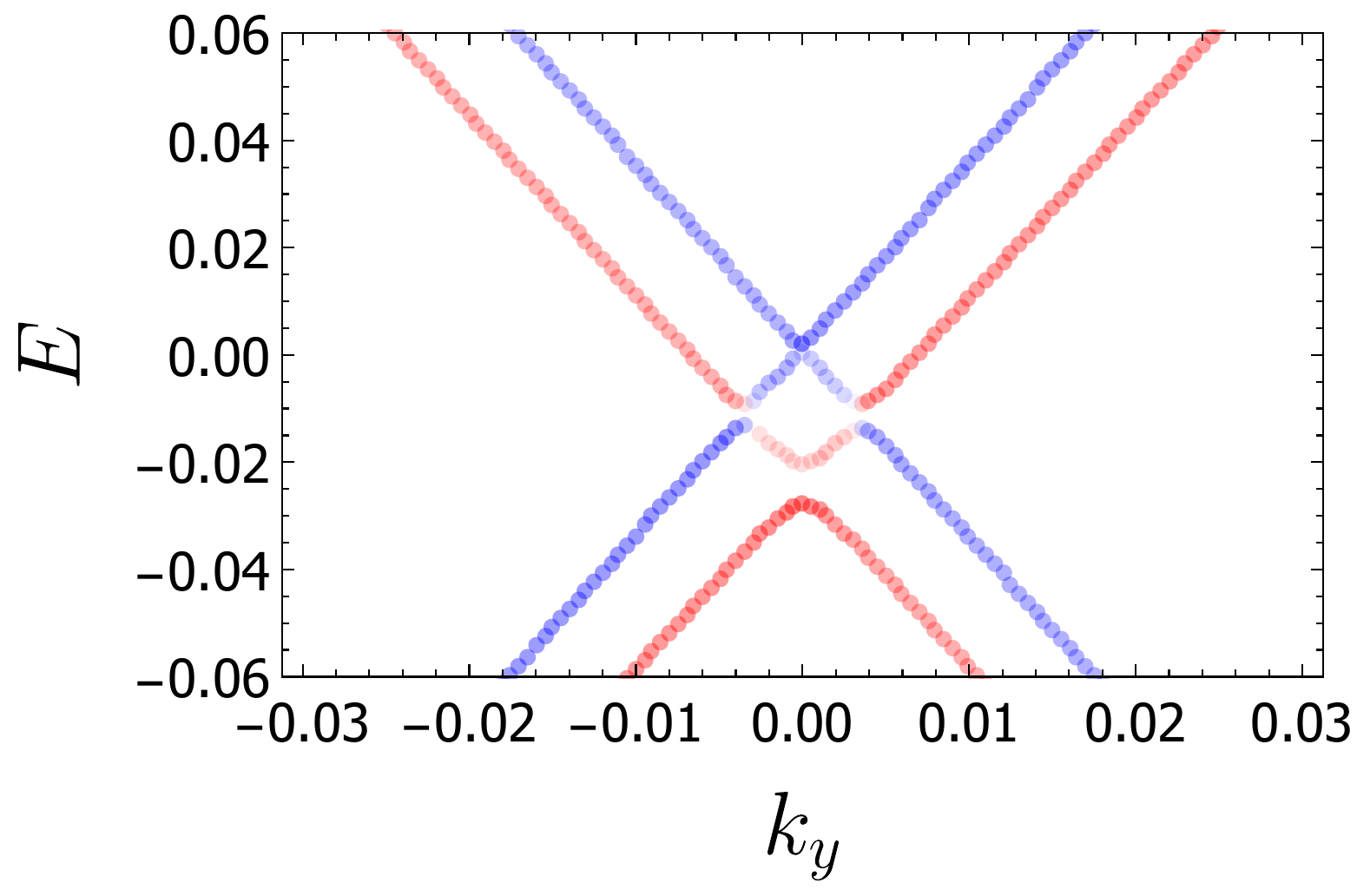}}
	\hspace{1em}
	\subfloat[\label{c3break4}]{%
	\includegraphics[width=0.3\linewidth]{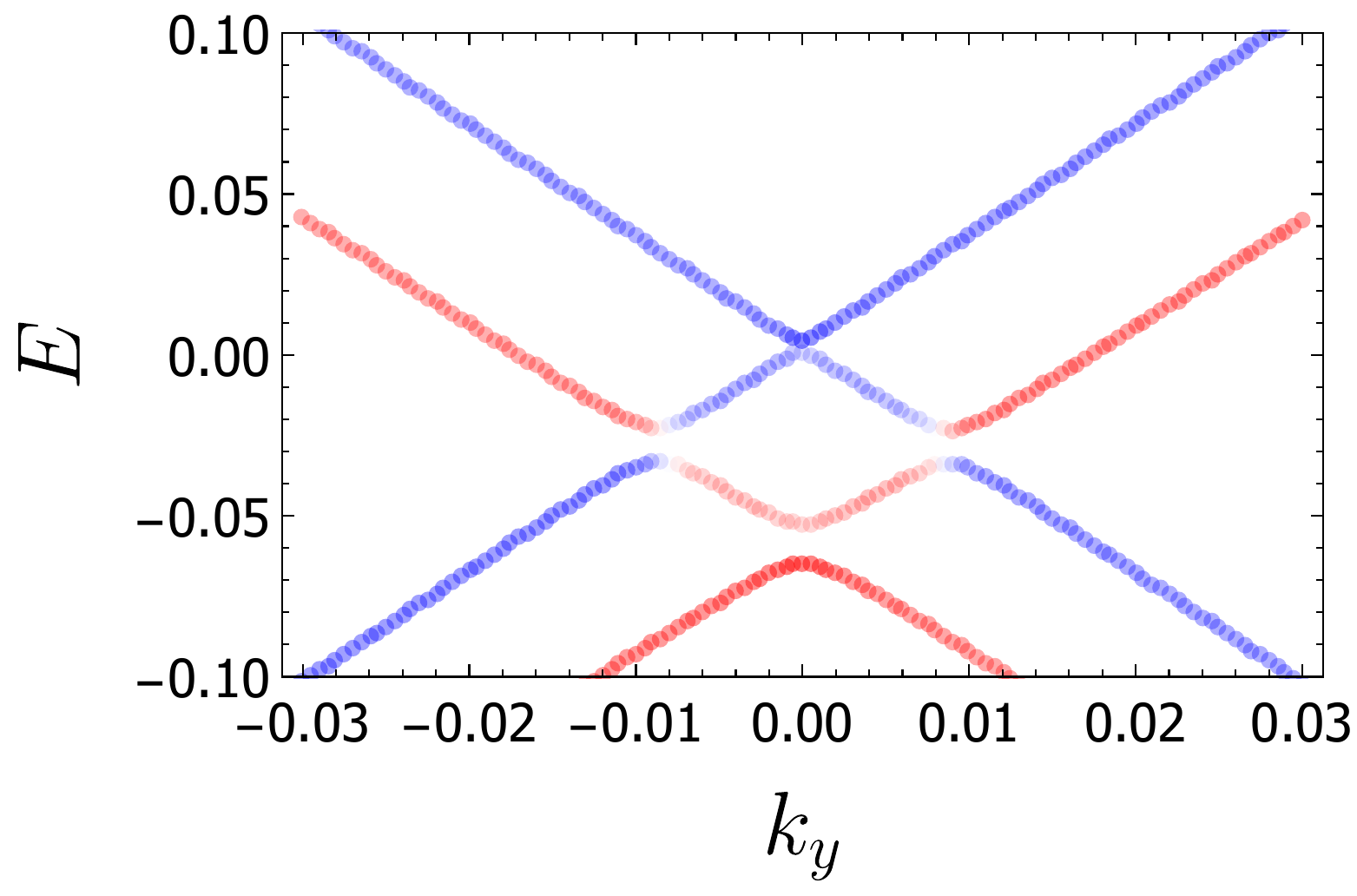}}
	\\
	\caption{\label{fig1} General interface between graphene and the WSM. (a) Displacement of graphene and the WSM lattices lead to a structure with no $C_3$ symmetry present. (b),(c) A zoom-in on the $\bf{K}'$ valley band structure with the tunneling conserving spin. (d),(e) A zoom-in on the $\bf{K}'$ valley band structure with spin non-conserving processes present. The colors code the average of the spin operator's $z$ component with red being $\uparrow$ and blue being $\downarrow$. The dilution of the colors in (d),(e) implies less spin polarization. Values of the coupling in (b),(d) are $\kappa_{A,0}=\kappa_{B,2}=\kappa_{B,3}=0.2$, $\kappa_{A,2}=\kappa_{A,3}=0.15$ and $\kappa_{B,0}=0.1$. Values of the coupling in (c),(e) are $\kappa_{A,0}=\kappa_{B,2}=\kappa_{B,3}=0.3$, $\kappa_{A,2}=\kappa_{A,3}=0.25$ and $\kappa_{B,0}=0.2$.}
\end{figure}

Having treated a highly symmetric model of hopping between the MLG and
the top layer of the WSM in the main text (see Fig. 3 in the main
text), we now treat the case where the $C_3$ symmetry is broken by the
coupling. We also allow the spin to not be conserved in
the tunneling process. We displace the two lattices with respect to
each other to get a configuration where there is no center of rotation
present (See Fig. \ref{fig1}\subref{str}). There are 6 atoms per unit
cell (at the interface) in our model; two of them being the A and the B sublattices in
graphene and the remaining four belong to the WSM which we denote by
numbers from 0 to 3. The coupling will take the general form
\begin{equation}
    \begin{pmatrix}
       \kappa_{A,0}(\vec{p}) & \kappa_{B,0}(\vec{p})\\
       \kappa_{A,1}(\vec{p}) & \kappa_{B,1}(\vec{p})\\
       \kappa_{A,2}(\vec{p}) & \kappa_{B,2}(\vec{p})\\
       \kappa_{A,3}(\vec{p}) & \kappa_{B,3}(\vec{p})\\
    \end{pmatrix}\otimes
    \begin{pmatrix}
    \nu_{\upuparrows} & \nu_{\uparrow\downarrow}\\
    \nu_{\downarrow\uparrow} & \nu_{\downdownarrows}
    \end{pmatrix},
\end{equation}    
where $\kappa_{A/B,i}(\vec{p})$ denotes the coupling between the A/B
sublattice and the $i$ site of the WSM which is generally a function
of momentum $\vec{p}$. $\nu_{\alpha\beta}$ denotes the coupling
between spins $\alpha$ and $\beta$ in graphene and the WSM,
respectively. Taking the configuration shown in Fig. \ref{fig1}\subref{str} and keeping nearest neighbor terms only, (1.1) will take the form, \begin{equation}
    \begin{pmatrix}
       \kappa_{A,0} & \kappa_{B,0}\\
       0 & 0\\
       \kappa_{A,2}e^{i\vec{p}.\vec{b_2}} & \kappa_{B,2}\\
       \kappa_{A,3}e^{i\vec{p}.\vec{b_1}+i\vec{p}.\vec{b_2}} & \kappa_{B,3}\\
    \end{pmatrix}\otimes
    \begin{pmatrix}
    \nu_{\upuparrows} & \nu_{\uparrow\downarrow}\\
    \nu_{\downarrow\uparrow} & \nu_{\downdownarrows},
    \end{pmatrix}
\end{equation}  
where $\vec{b_{1,2}}$ are the lattice bases vectors. The behavior of states in the ${\bf K}'$ valley depends
on the details of the coupling, as shown in
Fig. \ref{fig1}\subref{c3break1},\subref{c3break2}.
No qualitative difference from the highly symmetric model of the main
text is found in the spectrum near the ${\bf K}'$ valley. We conclude
that the spectrum presented in Fig. 4(c) of the main text is robust to
making the tunneling matrix elements between the MLG and the WSM
generic.
\section{Dissolution of Graphene States into Weyl Semimetal}
\label{SM:sec2}
It is well-known that unless there are symmetry restrictions, or
couplings are fine-tuned \cite{BIC-review}, a localized state which is
tunnel-coupled to a continuum of delocalized states will generically
hyrbridize with them and dissolve into them \cite{Mah00}. Here we
present numerical evidence that there is no hidden
symmetry/fine-tuning in our model that prevents this dissolution.

\begin{figure}[h]
	\subfloat[\label{amp}]{%
		\includegraphics[width=0.3\linewidth]{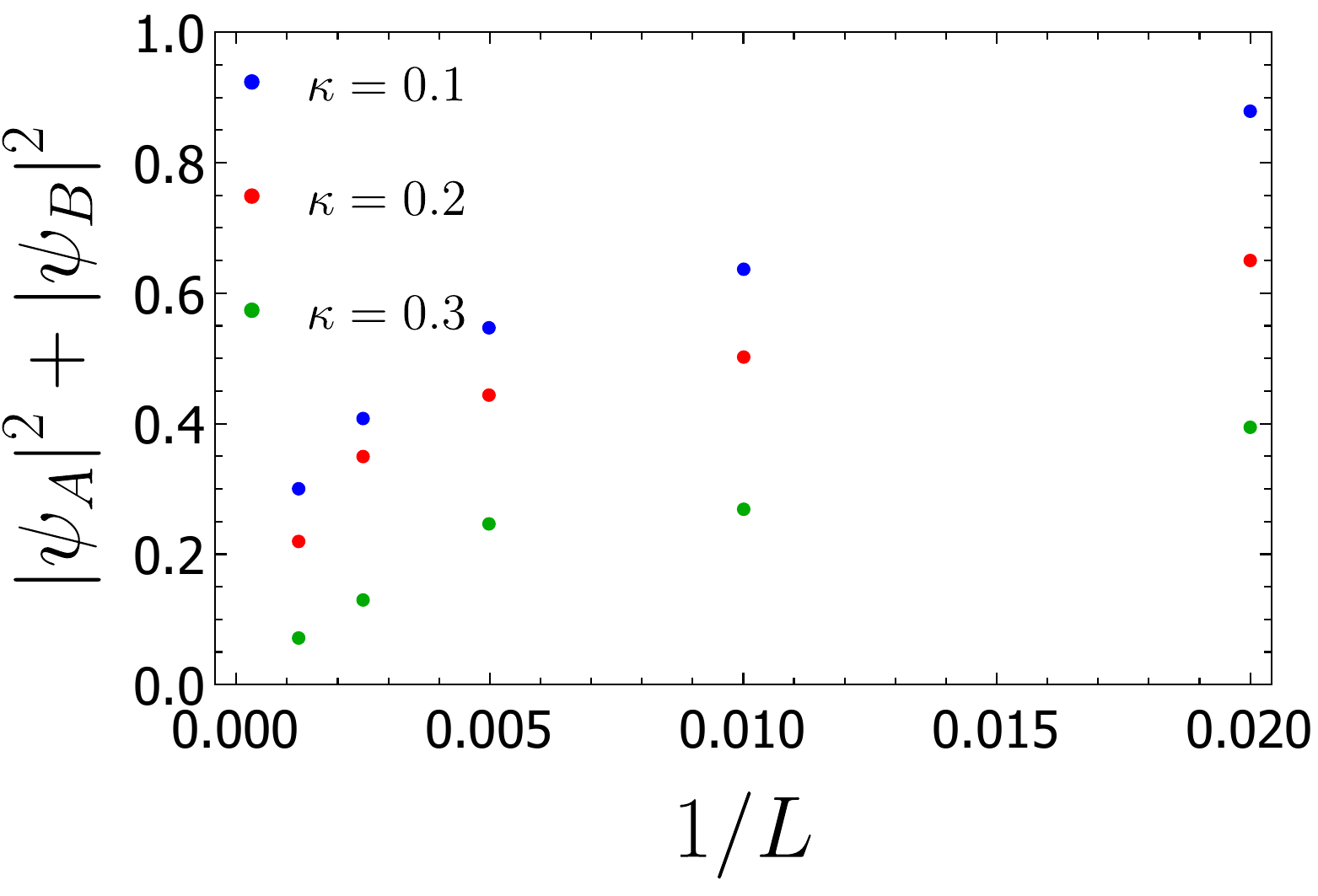}}
	\hspace{1em}
	%\subfloat[\label{ipr}]{%
	%	\includegraphics[width=0.4\linewidth]{ipr}}
	%	\hspace{1em}
		\subfloat[\label{amp2}]{%
		\includegraphics[width=0.3\linewidth]{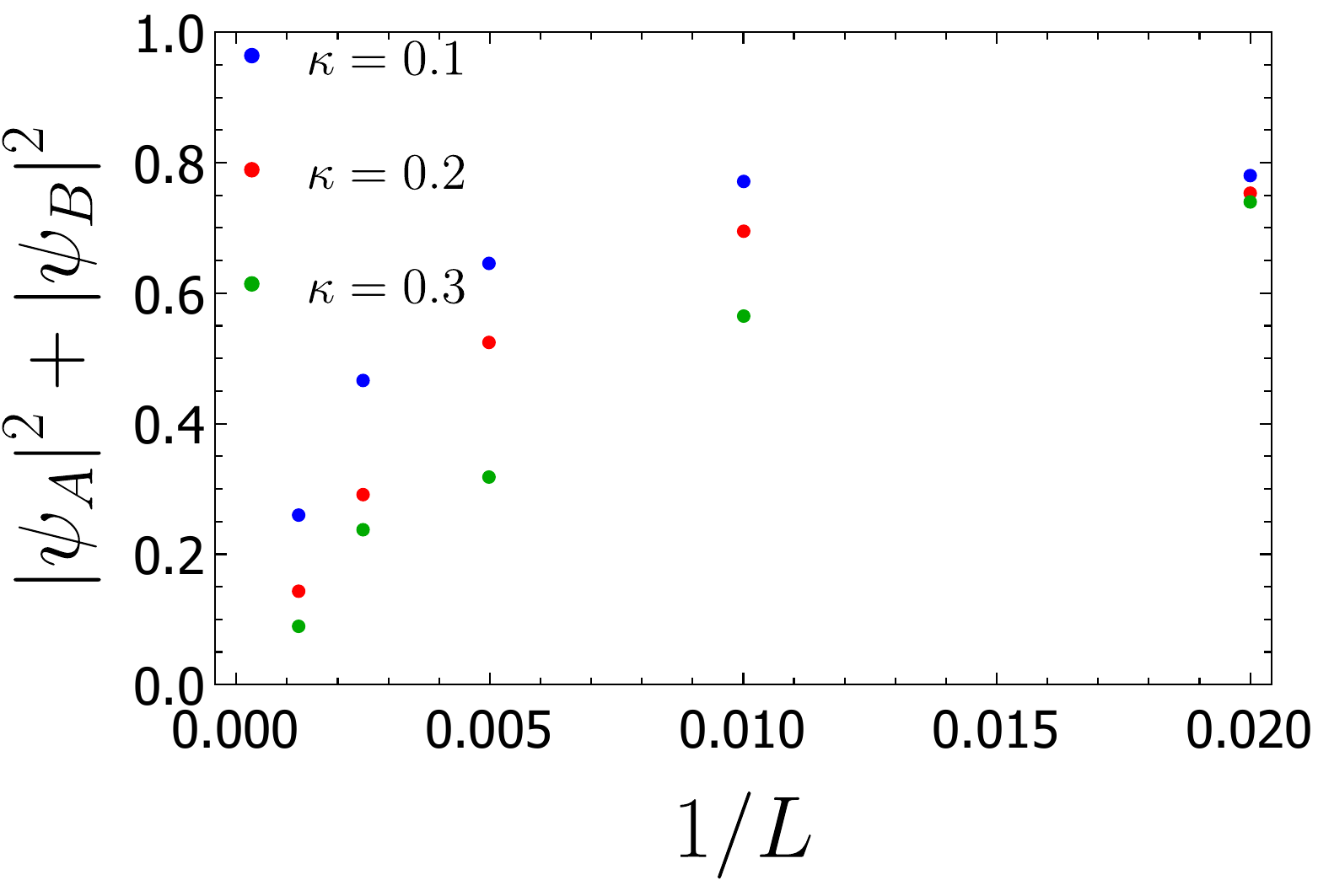}}
		\hspace{1em}
		\subfloat[\label{amp3}]{%
		\includegraphics[width=0.3\linewidth]{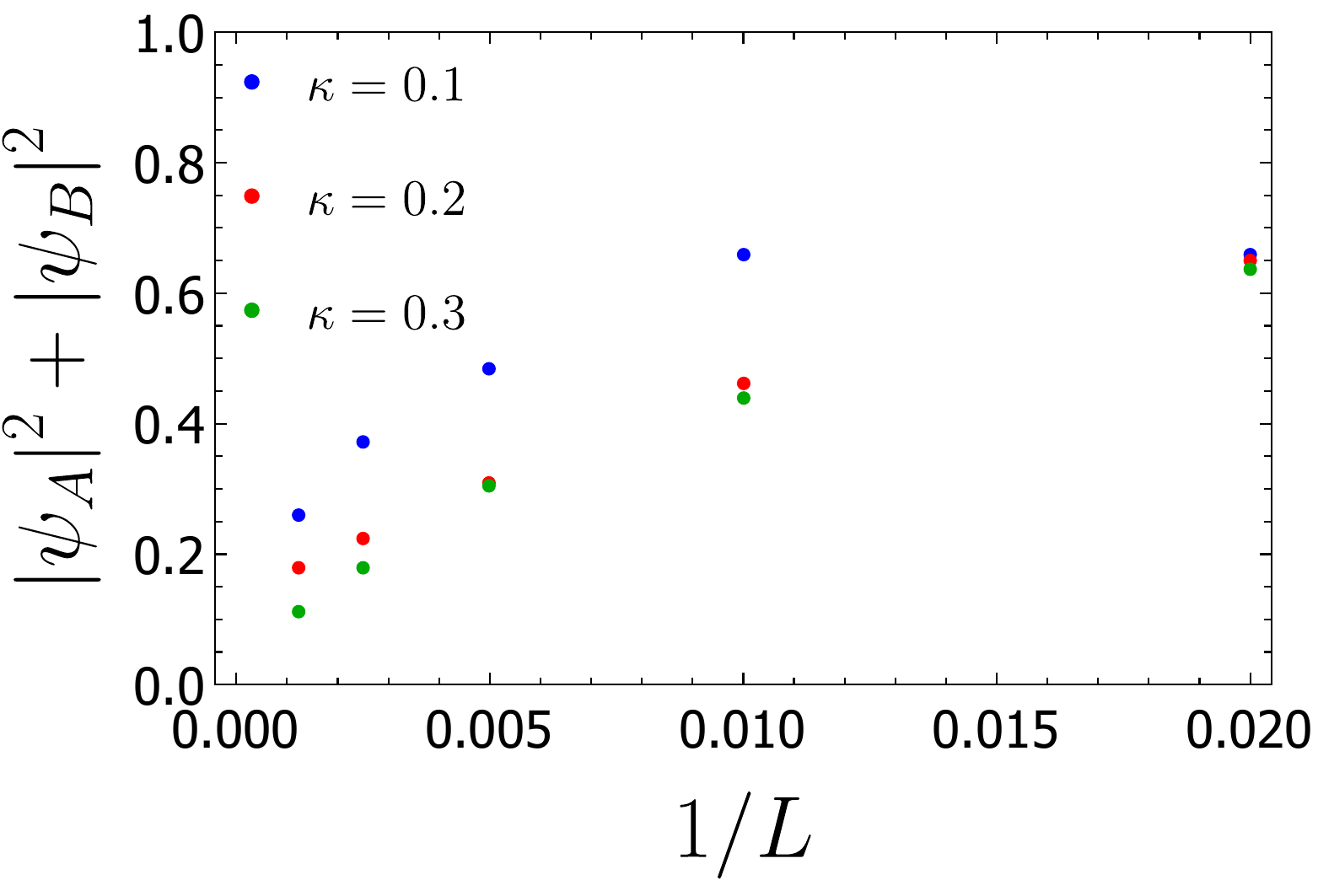}}\\
	\caption{\label{fig2}Delocalization of the $\bf{K}$ valley. (a)-(c) The graphene wave function amplitudes as a function of the system size for momenta $\vec{p}=(2\pi/3,0)$, $\vec{p}=(2\pi/3+0.025,0.014)$ and $\vec{p}=(2\pi/3-0.02,0.012)$ respectively. Calculations were done in the general configuration with the coupling in equation (1.2). $\kappa=0.1$ represents the set of parameters $\kappa_{A,0}=\kappa_{B,3}=0.1$ and $\kappa_{A,2}=\kappa_{A,3}=\kappa_{B,0}=\kappa_{B,2}=0.08$. $\kappa=0.2$ represents the set of parameters $\kappa_{A,0}=\kappa_{B,3}=0.2$ and $\kappa_{A,2}=\kappa_{A,3}=\kappa_{B,0}=\kappa_{B,2}=0.18$. $\kappa=0.3$ represents the set of parameters $\kappa_{A,0}=\kappa_{B,3}=0.3$ and $\kappa_{A,2}=\kappa_{A,3}=\kappa_{B,0}=\kappa_{B,2}=0.28$}
\end{figure}

To find the effect on a graphene state after the coupling is turned
on, for a given value of ${\vec k}$ near the ${\bf K}$ valley, we look
for states localized near the top surface, in a range of energies near
that of the graphene unperturbed states. We use the inverse
participation ratio ($IPR$) as a measure for localization which is
defined for a wave function $\psi$ as, \begin{equation}
  IPR=\frac{1}{\sum\limits_{i}\abs{\psi_i}^4}.
\end{equation}
where $i$ is a joint index representing the layer, spin, and for
graphene sites, sublattice as well.  A completely localized state will
give an $IPR$ of order 1 whereas a completely delocalized state will
give an $IPR$ of order $N$, with $N$ being the dimension of the
Hilbert space. Using this definition, we conduct a search for
localized states at a momentum near the ${\bf K}$ valley and in the
vicinity of the unperturbed energy. We then observe how the probability of the electron to be on the MLG
behaves as the
thickness of the WSM slab is increased and the bulk states become denser (See Fig. \ref{amp}). For
different momenta near the ${\bf K}$ valley, one sees clearly that the
states become delocalized as the the system size increases. One clearly sees the trend towards complete delocalization in the thermodynamic limit.
\section{Conductance}
\label{SM:sec3}
  We next discuss how one calculates the conductance in a graphene-WSM device \cite{beenakker}, for which the result is given in Fig. 4(d) in the main text. We work within the effective model for the $\bf{K}'$ valley introduced in equation (6) in the main text. In this simple model, there are four spin polarized bands and the spin carried by the bands alternates between up and down so it emulates the band inversion seen in the bands obtained from the tight binding calculation. In this model, however, the two middle bands touch at zero energy where there is a tiny gap in the bands obtained numerically. So as long as the chemical potential is away from this point (which is the regime we are interested in), the model should give qualitatively the same result as the numerical bands.  

\begin{figure}[h]
	\subfloat[\label{bands}]{%
		\includegraphics[width=0.4\linewidth]{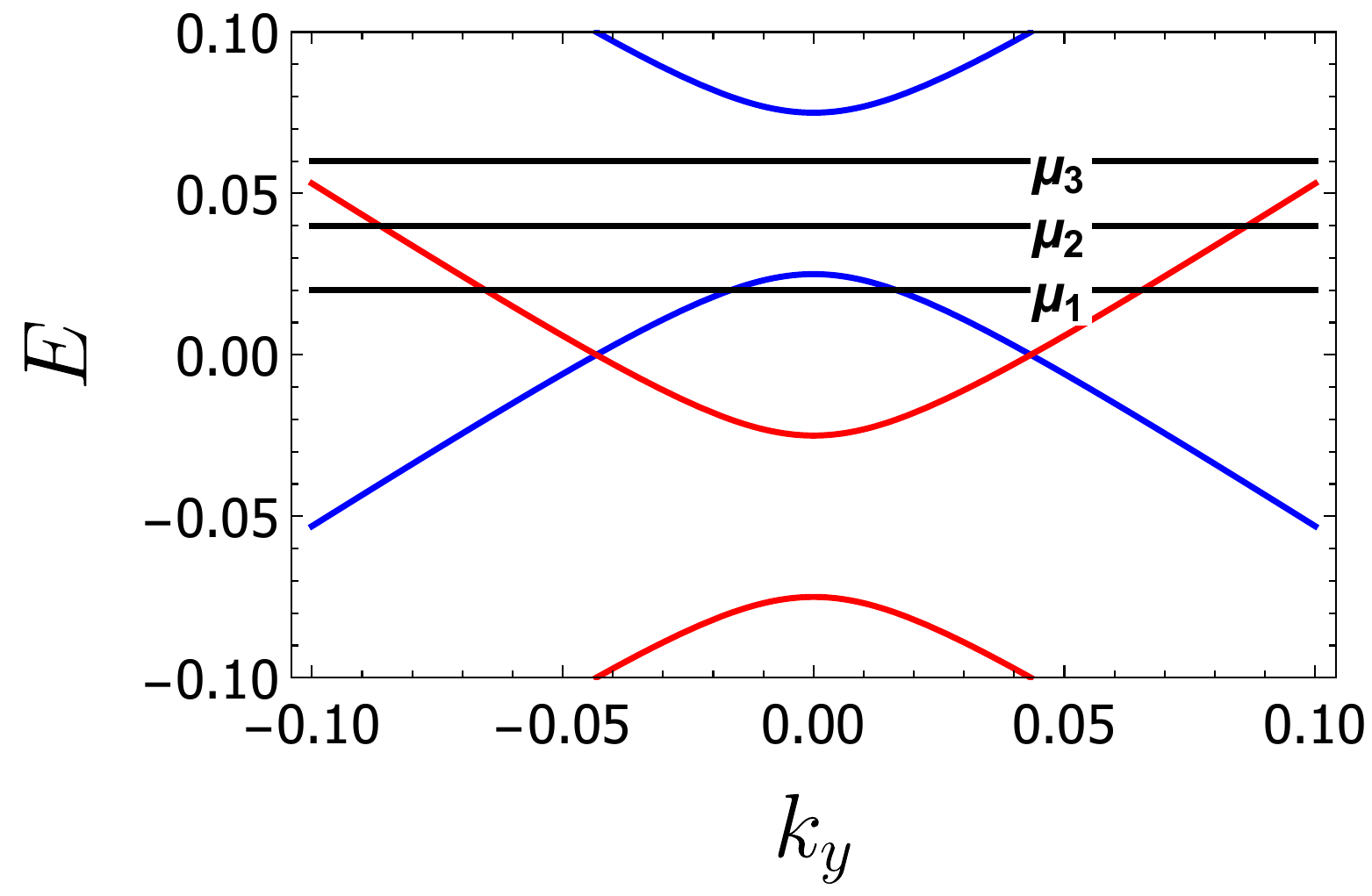}}
	\hspace{1em}
	\subfloat[\label{con}]{%
		\includegraphics[width=0.4\linewidth]{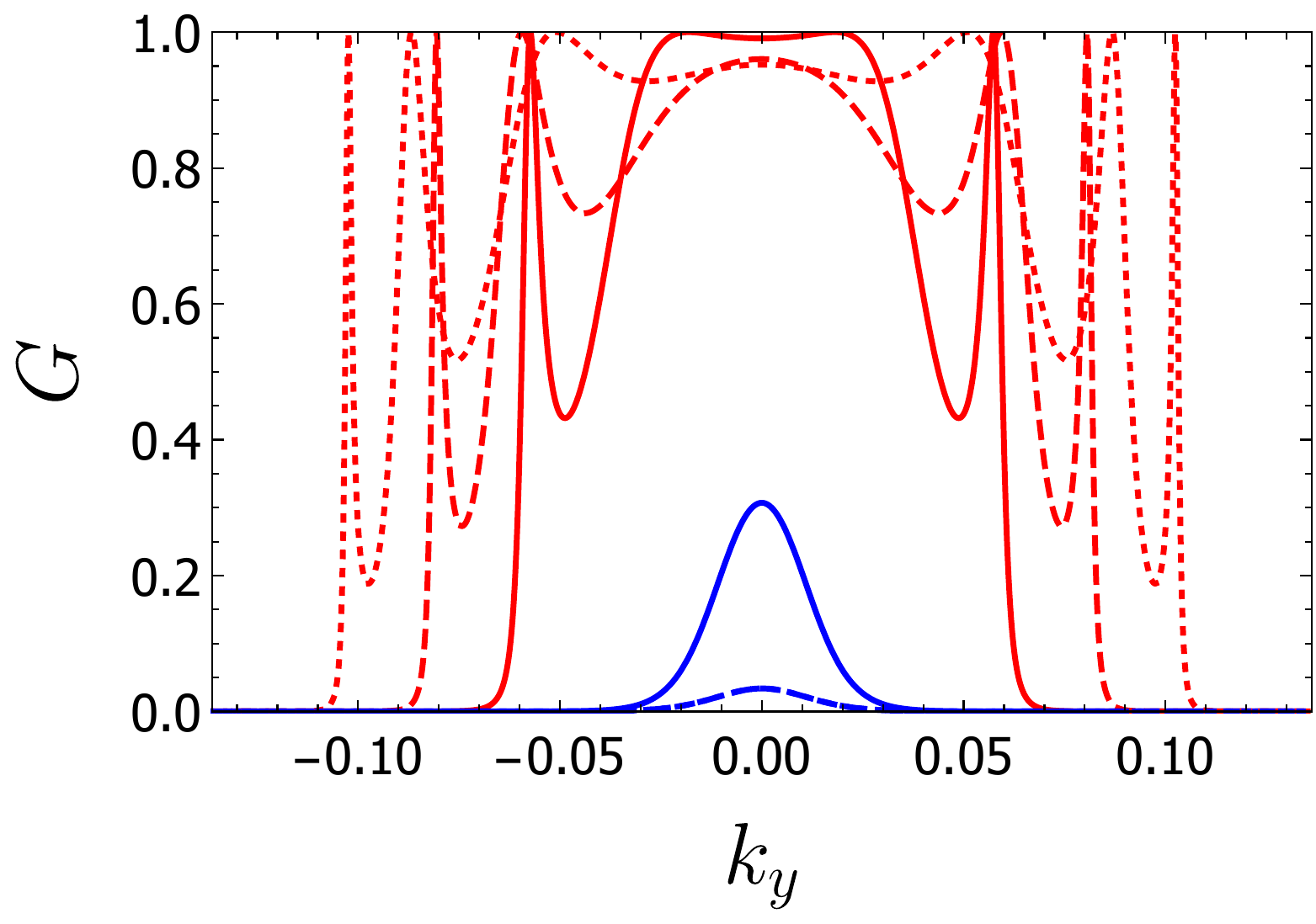}}\\
	\caption{\label{fig3}Conductance at different chemical potentials. (a) The bands obtained from the continuum model of equation (6) in the main text. The three horizontal black lines represent three values for the chemical potential. (b) The conductance measured in units of ($e^2/h$) for the three respective chemical potential. Solid, $\mu_1=0.02$, Dashed, $\mu_2=0.04$ and Dotted, $\mu_3=0.06$. In (a),(b) red denotes spin $\uparrow$ and blue denotes spin $\downarrow$.}
\end{figure}

Looking at the schematic picture in Fig. 1 in the main
text, we see that the strip is divided into three regions with
graphene leads on the left ($x<0$) and right ($x>L$) at potential 0
while the middle region ($0<x<L$) at potential $V_g$. We consider a
scattering problem where we assume the potential to take the
form \begin{equation} V(x) = \begin{cases} 0 & x\leq 0 \\ eV_g & 0\leq
    x\leq L \\ 0 & L\leq x
       \end{cases}\; .
    \end{equation}   
   For the ease of notation we denote the x-momentum $k$ and the transverse momentum $q$. We also set $\hbar$ and graphene Fermi velocity $v$ to 1. We show the conductance resulting from three different values for the chemical potential in Fig. \ref{fig3} to show that the result obtained is not too sensitive to the chemical potential. We present the detailed calculation for the chemical potential $\mu_1$ in Fig. \ref{fig3}; the others can be obtained in a similar fashion.\\ To start, we need to find the dispersion of electrons in the leads and in the middle region. We have in the leads \begin{equation}
       \varepsilon=\sqrt{k^2+q^2}
   \end{equation}   
For $0<x<L$, we have (using the continuum model of Eq. (6) in the main text)\begin{equation}
    \varepsilon=eV_g-\lambda_2+\sqrt{\lambda_1^2+\tilde{k}^2+q^2},
\end{equation}
for the spin up band. For the spin down band, we have \begin{equation}
    \varepsilon=eV_g+\lambda_2-\sqrt{\lambda_1^2+\tilde{k}^2+q^2}
\end{equation}
Note that $k$ in the leads is always real while $\tilde{k}$ can be imaginary.

Consider a scattering state coming from the left. We have for $x<0$, %\begin{equation}
        \begin{align}
    \Phi_L &= \begin{pmatrix}
           -1 \\
           z_k 
         \end{pmatrix}e^{ikx+iqy}+r\begin{pmatrix}
           -1 \\
           z_{-k} 
         \end{pmatrix}e^{-ikx+iqy}
        \end{align}
       %\end{equation}
For $0<x<L$, we have
     \begin{align}
       \tilde{\Phi} = \alpha\begin{pmatrix}
           -1 \\
           w_{\tilde{k}} 
         \end{pmatrix}e^{i\tilde{k}x+iqy}+\beta\begin{pmatrix}
           -1 \\
           w_{-\tilde{k}} 
         \end{pmatrix}e^{-i\tilde{k}x+iqy}
      \end{align}
For $x>L$, we have %\begin{equation}
    \begin{align}
    \Phi_R &= t\begin{pmatrix}
           -1 \\
           z_k 
         \end{pmatrix}e^{ik(x-L)+iqy}
  \end{align}
%\end{equation}
$z_k=\frac{k-iq}{\sqrt{k^2+q^2}}$. $w_{\tilde{k}}=\frac{\tilde{k}-iq}{\lambda_1+\sqrt{\lambda_1^2+\tilde{k}^2+q^2}}$
for the spin up states, and
$w_{\tilde{k}}=\frac{\tilde{k}-iq}{\lambda_1-\sqrt{\lambda_1^2+\tilde{k}^2+q^2}}$
for the spin down states. $r$ and $t$ define the reflection and the
transmission probabilities, respectively.\\ Using the continuity of
the wave functions at $x=0$ and $x=L$, we can solve for the
transmission amplitude $t$, resulting in \begin{equation}
  t=\frac{(1+z_k^2)(w_{\tilde{k}}-w_{-\tilde{k}})}{e^{i\tilde{k}L}[(1+z_{k}w_{-\tilde{k}})(w_{\tilde{k}}-z_{k})]+e^{-i\tilde{k}L}[(1+z_{k}w_{\tilde{k}})(z_{k}-w_{-\tilde{k}})]}
\end{equation}
The transmission probability is given by $T=|t|^2$. Plotting this
equation as a function of the conserved transverse momentum ($k_y$)
gives Fig. (4)(d) in the main text.
\section{Graphene-Fermi arc hybridization}
\label{SM:sec4}
\begin{figure}[h]
	\centering
	\includegraphics[width=0.3\linewidth]{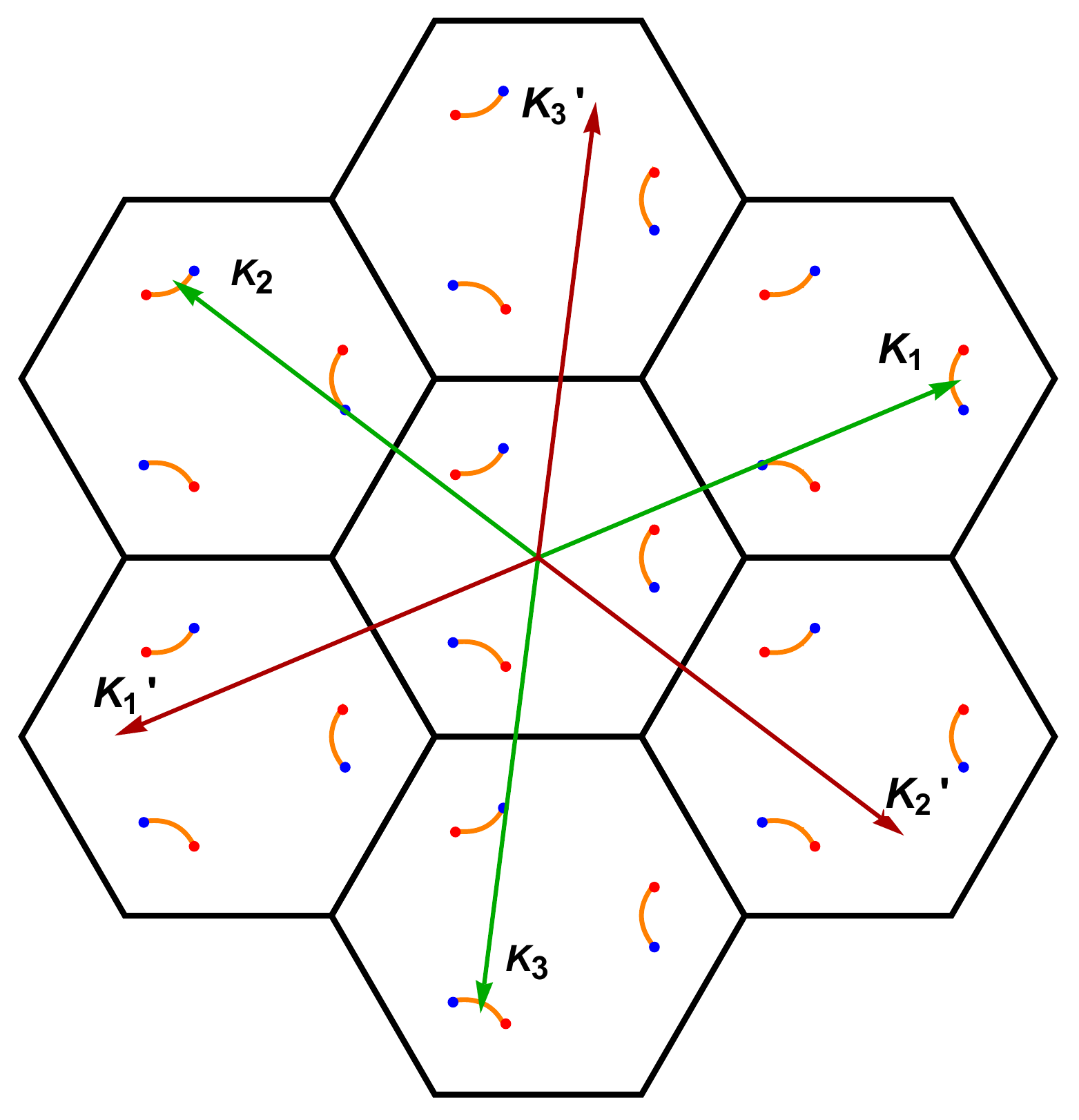}
	\caption{an incommensurate structure of graphene and the WSM shown in momentum space. The 3 equivalent graphene $\bf{K}$ vectors coincide with the Fermi arcs.}
	\label{fig:bz2}
\end{figure} 
Another possibility that this system offers is the hybridization between graphene's Dirac cones and the Fermi arc states localized on the surface of the WSM. This again will occur in only one of graphene's valleys due to the breaking of time reversal symmetry. This construction, however, requires some fine-tuning as we need the chemical potential to be at the Dirac points in graphene and at the Weyl nodes in the WSM. Furthermore, we need the $K$ point in graphene to coincide with the Fermi arcs as shown in Fig. \ref{fig:bz2} which happens at certain orientations of the graphene-WSM system.\\ 

\begin{figure}[h]
	%\centerfig
	\subfloat[\label{kbands}]{%
		\includegraphics[width=0.3\linewidth]{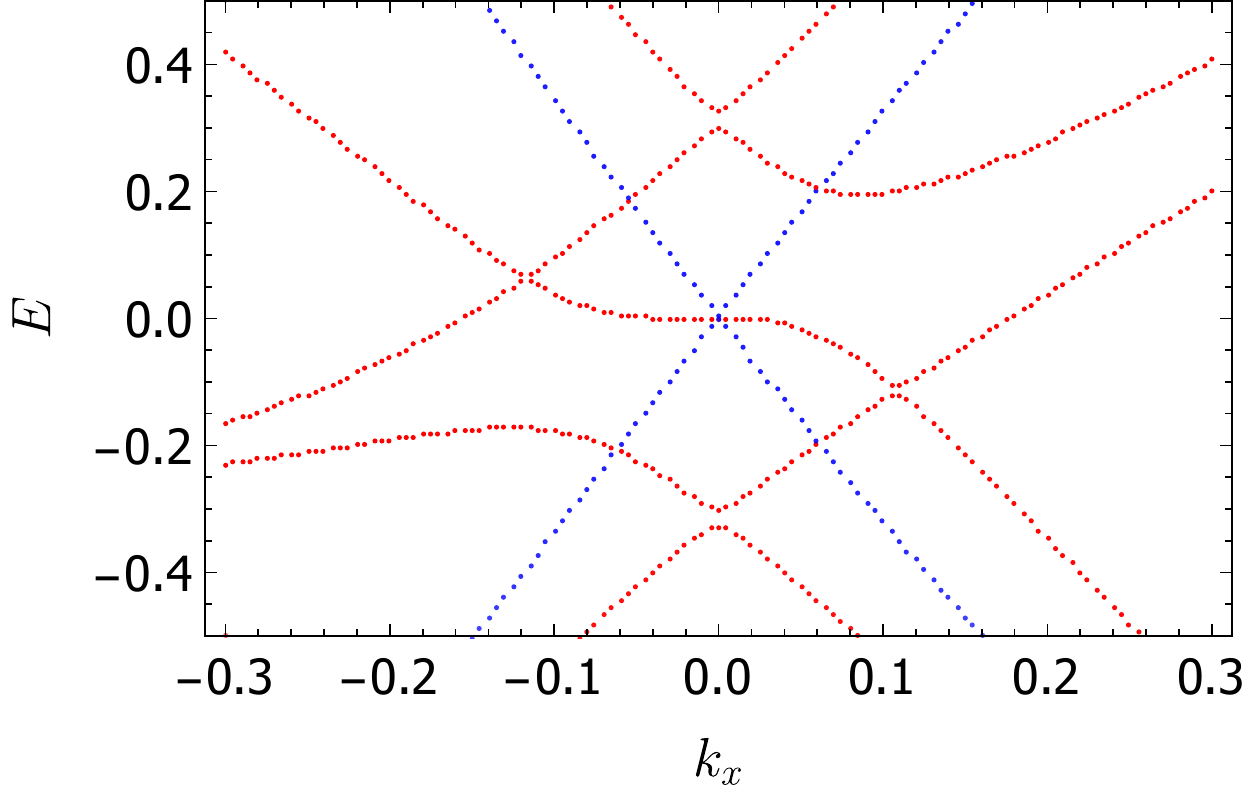}}
	%\hspace{\fill}
	\subfloat[\label{recon}]{%
		\includegraphics[width=0.3\linewidth]{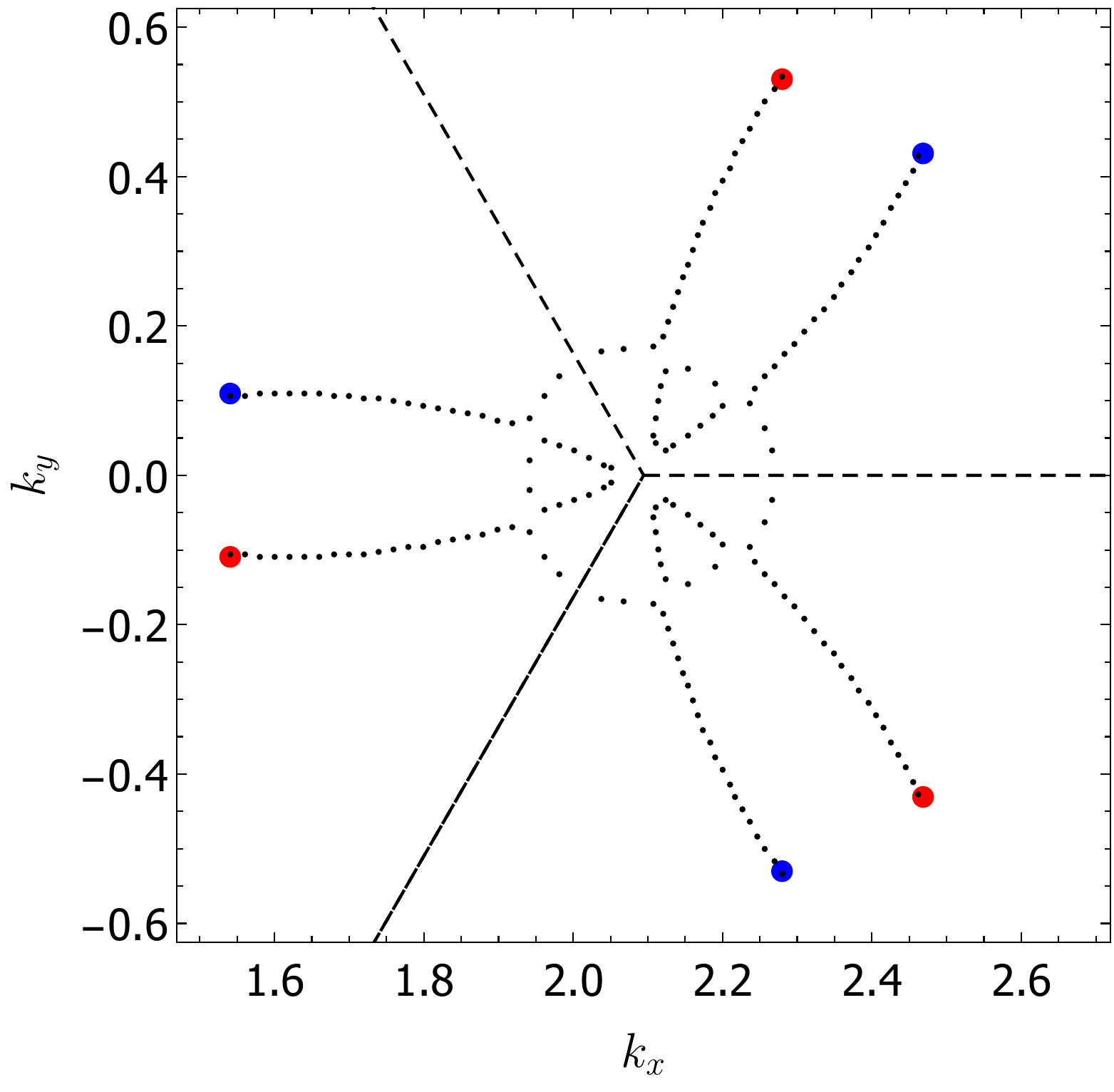}}\\
		\subfloat[\label{linear}]{%
		\includegraphics[width=0.3\linewidth]{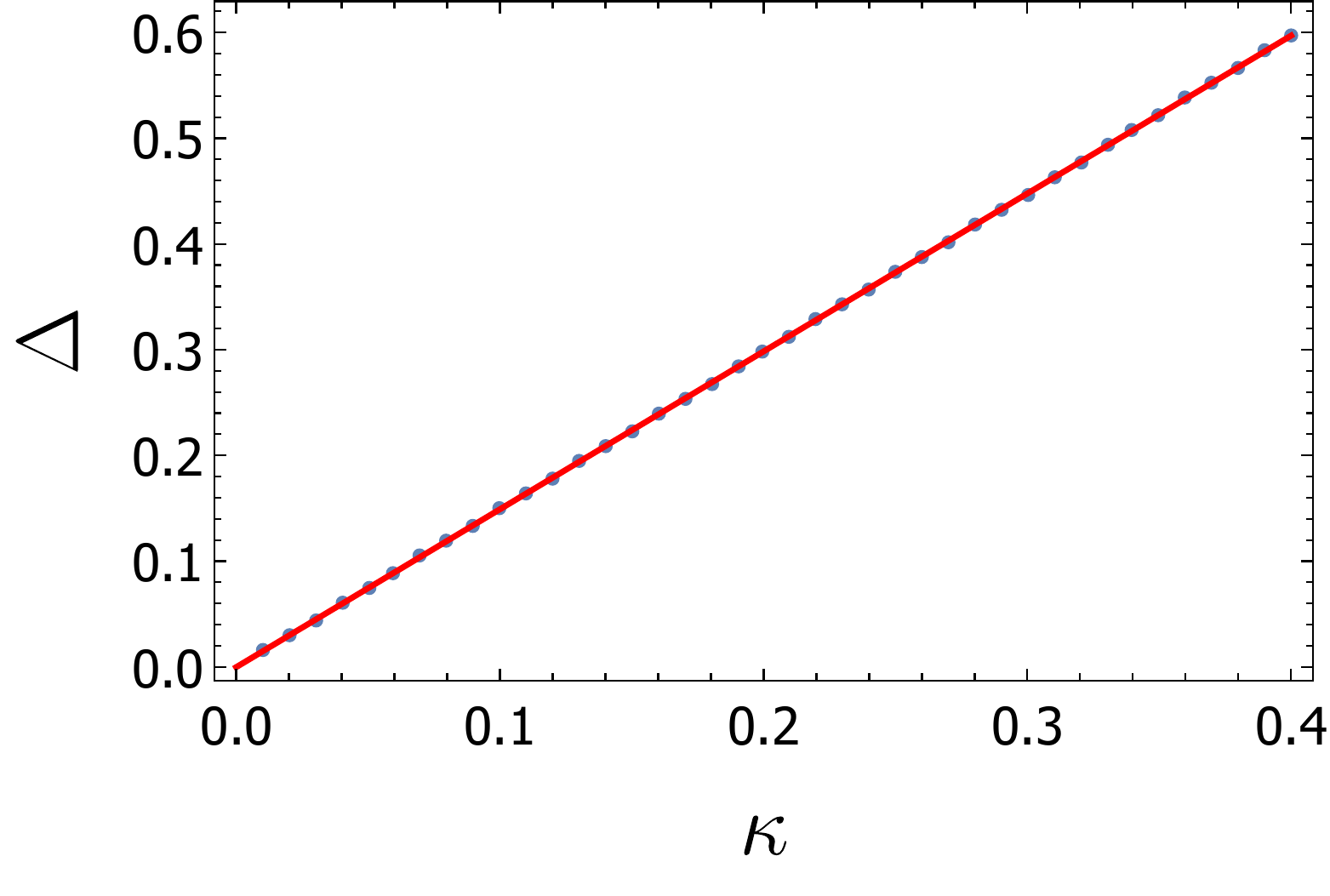}}
		\subfloat[\label{quadratic}]{%
		\includegraphics[width=0.3\linewidth]{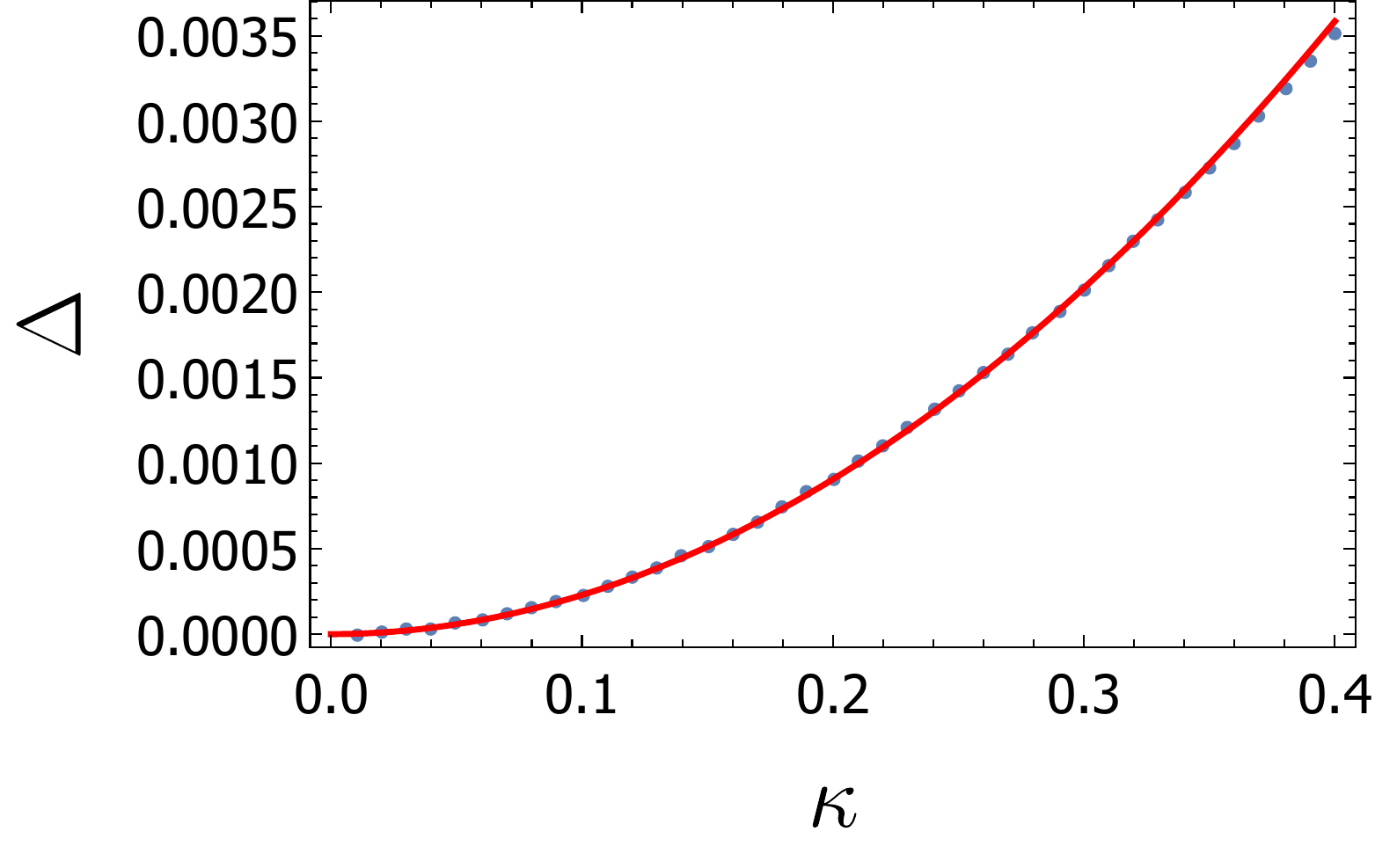}}\\
	\caption{\label{fig5} Fermi Arcs hybridization with the $\bf{K}$ valley. (a) The Dirac cone near the $\bf{K}$ valley along the $k_x$ axis in momentum hybridizing with Fermi arc states present on the top surface of the WSM. (b) A plot of the reconstructed Fermi arcs near the $\bf{K}$ point which shows the detachement of the arcs into three closed loops. (c) The gap in the spin $\uparrow$ Dirac cone scales linearly with the coupling as the data points fits nicely to a straight line (in red). (d) The gap in the spin $\downarrow$ Dirac cone scales quadratically with the coupling as the data is fitted nicely to a parabola (in red).}
\end{figure}

In a fashion similar to our treatment in the main text, we consider a commensurate structure of graphene and the WSM with the Dirac $\bf{K}$ point overlapping a Fermi arc in momentum space. We then perform our tight binding calculations to study the electronic properties of this system. A plot for the band structure near the $\bf{K}$ valley shows the effect of the WSM Fermi arc on the Dirac cone is shown in Fig. \ref{fig5}\subref{kbands}.  

We can understand the resulting band structure based on the fact that the Fermi arc states are spin polarized in our simple model with the spins being up for the top WSM surface. Consequently, the spin up Dirac cone acquires a gap which is shown to scale linearly with the coupling as shown in Fig. \ref{fig5}\subref{linear}. The spin down cone, however, remains relatively intact (See Fig. \ref{fig5}\subref{kbands}) but upon careful examination, we can see it acquires a small gap which scales quadratically with the coupling as shown in Fig. \ref{fig5}\subref{quadratic}.\\

Finally, looking at the zero energy contour, we can see a reconstruction of the Fermi arcs as shown in Fig. \ref{fig5}\subref{recon}. 
The existence of a Fermi loop at the surface can lead to interesting resonance signatures in the optical conductivity of the system in a magnetic field perpendicular to the surface.